# ECT* PREPRINT

# Model–space approach to $^1S_0$ neutron and proton pairing with the Bonn meson–exchange potentials

Ø. Elgarøy[a], L. Engvik[a], M. Hjorth–Jensen[b] and E. Osnes[a]

[a]Department of Physics, University of Oslo, N-0316 Oslo, Norway

[b]ECT*, European Centre for Theoretical Studies in Nuclear Physics and Related Areas, Trento, Italy

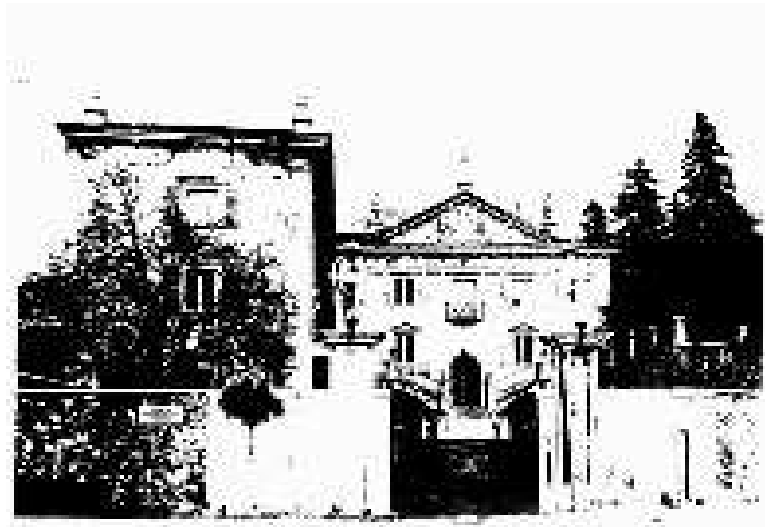

# Model–space approach to $^1S_0$ neutron and proton pairing with the Bonn meson–exchange potentials

Ø. Elgarøy, L. Engvik

*Department of Physics, University of Oslo, N–0316 Oslo, Norway*

M. Hjorth–Jensen,

*ECT\*, European Centre for Theoretical Studies in Nuclear Physics and Related Areas, Strada delle Tabarelle 286, I-38050 Villazzano (Trento), Italy*

E. Osnes

*Department of Physics, University of Oslo, N–0316 Oslo, Norway*

In this work we calculate neutron and proton energy gaps in neutron star matter, using the Bonn meson–exchange interactions and a model–space approach to the gap equation. This approach allows a consistent calculation of energy gaps and single particle energies with the model–space Brueckner–Hartree–Fock (MBHF) method, without double counting of two–particle correlations. Neutron energy gaps are calculated at zero and finite temperature. Proton energy gaps are calculated at beta equilibrium, and it is shown that the inclusion of muons has a significant effect. The results are compared with those of other works, and the implications for neutron star physics are briefly discussed.

## I. INTRODUCTION

Superfluidity in nuclear matter has over the last 30 years received a great deal of attention, partly due to the possible impact on neutron star physics [1]. The existence of superfluid neutrons in the inner crust, and superfluid neutrons together with superconducting protons in the quantum liquid region of neutron stars must be considered as well established. Among the physical consequences of this qualitative picture, the most interesting is probably the role played by hadron superfluidity in cooling processes and in explaining pulsar glitches [2]. However, to assess the quantitative impact of superfluidity, precise knowledge of the size of the pairing energy gap is needed. Obtaining accurate results for this quantity in neutron star matter has proved to be a difficult problem. The most overwhelming difficulty is probably the consistent inclusion of medium effects in the effective pairing interaction, but even at the two–body level the results obtained so far are disturbingly scattered. This work is an attempt to clarify the situation at the two–body level, following an approach to the BCS gap equation introduced by Anderson and Morel [3] and applied to pairing in nuclear matter by Baldo et al. [4]. This approach to the gap equation allows us to calculate single–particle energies in the model–space Brueckner–Hartree–Fock (MBHF) approach in the same model space in which the gap equation is solved. In our calculations, we employ the meson–exchange potential models of the Bonn group [5], where the only parameters entering are physically motivated ones, like coupling constants and vertex cutoffs. In applications to finite nuclei and nuclear matter [6,7], the Bonn potentials have given satisfactory results.

The structure of this work is as follows: In section 2 we reformulate the gap equation in terms of two coupled equations. We show how this reformulation allows us to renormalize the bare nucleon–nucleon (NN) interaction without introducing double counting of two–particle correlations in the gap equation. A brief description of the MBHF method is given, before we discuss the numerical solution of the gap equation. In section 3 we apply the method to pure neutron matter at zero and finite temperature, using the Bonn potentials of ref. [5]. In section 4 we investigate neutron star matter at $\beta$ equilibrium, taking into account the appearance of muons at higher densities. In sections 5 and 6 we compare our results to those of similar works, and discuss the influence of the choice of theoretical framework on the results.

## II. THEORY

When the attraction in the $^1S_0$ partial wave is dominant, BCS theory [8,9] predicts a transition to a superfluid(neutrons)/superconducting(protons) state if a non–zero solution $\Delta_k$ of the gap equation

$$\Delta_k = -\sum_{\mathbf{k}'} V_{k,k'} \frac{\Delta_{k'}}{2E_{k'}} \qquad (1)$$

can be found, where $\Delta_k$ is the so-called gap function. Here $\mathbf{k}, \mathbf{k}'$ are single–particle momenta[1], $V_{k,k'}$ is the $^1S_0$ component of the nucleon–nucleon interaction in momentum space, which is independent of the orientation of $\mathbf{k}$ and $\mathbf{k}'$, and $E_k = \sqrt{(\epsilon_k - \epsilon_{k_F})^2 + \Delta_k^2}$ is the quasiparticle energy, where $\epsilon_k$ and $\epsilon_{k_F}$ are the single–particle energies at $k$ and $k_F$ respectively. The so-called decoupling approximation is assumed, that is, we take the Fermi surface to be sharp even in the presence of pairing correlations. This was found in ref. [10] to be a good approximation. Due to the decoupling approximation, the determination of single–particle energies is left as a separate problem which we will discuss later in this paper.

The solution of equation (1) is complicated by the repulsive core of the nucleon–nucleon (NN) interaction, and a method for overcoming this difficulty has to be devised. One particular scheme is to replace the uncorrelated plane–wave basis used in the BCS approach with a basis in which short–range correlations are included, the so-called Correlated Basis Functions (CBF) approach, see Krotcheck and Clark [11] for an overview. Other authors have replaced the pairing matrix elements $V_{kk'}$ with the corresponding $G$–matrix elements [12,13], or with a pairing interaction constructed from Skyrme forces [14]. Another alternative was proposed in the work of Baldo *et al.* [4], where it was shown that the gap equation could be conveniently reformulated so that its role in accounting for short–range correlations becomes more transparent and the numerical treatment easier. As their description was rather brief, we will re–derive their result here.

We split the summation in the gap equation (1) into two parts by introducing an arbitrary cutoff $k_C$ in momentum space:

$$\Delta_k = -\sum_{k' \leq k_C} V_{kk'} \frac{\Delta_{k'}}{2E_{k'}} - \sum_{k' > k_C} V_{kk'} \frac{\Delta_{k'}}{2E_{k'}}. \qquad (2)$$

Equation (2) is now taken as basis for an iterative procedure. As a first approximation, $\Delta^{(0)}$, we neglect the second sum in (2) and take

$$\Delta_k^{(0)} = -\sum_{k' \leq k_C} V_{kk'} \frac{\Delta_{k'}}{2E_{k'}}.$$

Inserting this in the second sum in (2) gives a new approximation to the energy gap, $\Delta^{(1)}$:

$$\begin{aligned}
\Delta_k^{(1)} &= -\sum_{k' \leq k_C} V_{kk'} \frac{\Delta_{k'}}{2E_{k'}} + \sum_{k' > k_C} \sum_{k'' \leq k_C} V_{kk'} \frac{1}{2E_{k'}} V_{k'k''} \frac{1}{2E_{k''}} \Delta_{k''} \\
&= -\sum_{k' \leq k_C} V_{kk'} \frac{\Delta_{k'}}{2E_{k'}} + \sum_{k' \leq k_C} \sum_{k'' > k_C} V_{kk''} \frac{1}{2E_{k''}} V_{k''k'} \frac{1}{2E_{k'}} \Delta_{k'} \\
&= -\sum_{k' \leq k_C} (V_{kk'} - \sum_{k'' > k_C} V_{kk''} \frac{1}{2E_{k''}} V_{k''k'}) \frac{\Delta_{k'}}{2E_{k'}}.
\end{aligned}$$

We proceed by inserting the expression for $\Delta^{(1)}$ in the second sum in (2) to obtain a new approximation $\Delta^{(2)}$. As long as $\Delta_k$ in the first sum is taken to be the exact solution of the gap equation for $k \leq k_C$, we need only iterate in the second sum. Continued iteration of this procedure gives an exact expression for the energy gap:

$$\begin{aligned}
\Delta_k = -\sum_{k' \leq k_C} (V_{kk'} &- \sum_{k'' > k_C} V_{kk''} \frac{1}{2E_{k''}} V_{k''k'} \\
&+ \sum_{k'' > k_C} \sum_{k''' > k_C} V_{kk'''} \frac{1}{2E_{k'''}} V_{k'''k''} \frac{1}{2E_{k''}} V_{k''k'} + \cdots) \frac{\Delta_{k'}}{2E_{k'}}.
\end{aligned}$$

By defining

$$\begin{aligned}
\widetilde{V}_{kk'} = V_{kk'} &- \sum_{k'' > k_C} V_{kk''} \frac{1}{2E_{k''}} V_{k''k'} \qquad (3) \\
&+ \sum_{k'' > k_C} \sum_{k''' > k_C} V_{kk'''} \frac{1}{2E_{k'''}} V_{k'''k''} \frac{1}{2E_{k''}} V_{k''k'} - \cdots,
\end{aligned}$$

---

[1]Throughout this work we set $\hbar = c = 1$.

the gap equation becomes

$$\Delta_k = - \sum_{k' \leq k_C} \widetilde{V}_{kk'} \frac{\Delta_{k'}}{2E_{k'}}. \tag{4}$$

Furthermore, we have

$$\widetilde{V}_{kk'} = V_{kk'} - \sum_{k''>k_C} V_{kk''} \frac{1}{2E_{k''}} (V_{k''k'} - \sum_{k'''>k_C} V_{k''k'''} \frac{1}{2E_{k'''}} V_{k'''k'} + \cdots)$$

$$= V_{kk'} - \sum_{k''>k_C} V_{kk''} \frac{1}{2E_{k''}} \widetilde{V}_{k''k'} \tag{5}$$

This shows that an equivalent formulation of the gap equation (1) is given by the coupled integral equations

$$\widetilde{V}_{kk'} = V_{kk'} - \sum_{k''>k_C} V_{kk''} \frac{1}{2E_{k''}} \widetilde{V}_{k''k'} \tag{6}$$

$$\Delta_k = - \sum_{k' \leq k_C} \widetilde{V}_{kk'} \frac{1}{2E_{k'}} \Delta_{k'}, \tag{7}$$

which have to be solved simultaneously. We stress that there are no new approximations involved in these equations, and that they therefore represent a totally equivalent formulation of the gap equation (1). The role of the effective interaction $\widetilde{V}$ is to sum up all virtual transitions to momenta above $k_C$. The advantage of this procedure is twofold. First, through the renormalization in equation (6) we obtain a more well-behaved pairing interaction, because the repulsive core of the bare interaction is integrated out. Secondly, we can explicitly see that double counting of two-particle correlations is avoided. Excitations to intermediate states above $k_C$ are included in $\widetilde{V}$, while excitations to states below $k_C$ are included in the gap equation (7), see fig.1. Double counting will arise if the bare interaction $V$ is replaced with any sort of effective interaction, e.g. the Brueckner $G$-matrix, which contains excitations to states below $k_C$. This corresponds to taking $k_C = k_F$ in equation (6), while at the same time taking $k_C = \infty$ in equation (7), which is clearly inconsistent, see fig. 2. Thus caution is needed if pairing matrix elements are obtained from some sort of medium-renormalized interaction. One should then first find out which two-body correlations are included in the effective interaction, and then reformulate the gap equation so that the two-body correlations included in the effective interaction are excluded from the gap equation.

Our calculation of pairing gaps is a two-step process: First we solve self-consistently the Brueckner-Hartree-Fock (BHF) equations for the single-particle energies, using a $G$-matrix defined through the Bethe-Brueckner-Goldstone equation

$$G = V + V \frac{Q}{\omega - H_0} G. \tag{8}$$

Here $V$ is the NN interaction, $Q$ is the Pauli operator which prevents scattering into intermediate states prohibited by the Pauli principle, $H_0$ is the unperturbed Hamiltonian acting on the intermediate states and $\omega$ is the so-called starting energy, the unperturbed energy of the interacting nucleons. Methods to solve this equation are reviewed in ref. [6]. The single-particle energies for states $k_i$ ($i$ encompasses all relevant quantum numbers like momentum, isospin projection, angular momentum, spin etc.) in nuclear matter are assumed to have the simple quadratic form

$$\epsilon_{k_i} = \frac{k_i^2}{2m^*} + \delta_i, \tag{9}$$

where $m^*$ is the effective mass. The terms $m^*$ and $\delta$, the latter being an effective single-particle potential related to the $G$-matrix, are obtained through the self-consistent BHF-procedure. The so-called model-space BHF-method is used [6,16], with a cutoff $k_M > k_F$ in momentum space. In this approach the single-particle spectrum is defined by

$$\epsilon_{k_i} = \frac{k_i^2}{2m} + u_i \tag{10}$$

where $m$ is the nucleon mass, and the single-particle potential $u_i$ is given by

$$u_i = \sum_{k_h \leq k_F} \langle k_i k_h | G(\omega = \epsilon_{k_i} + \epsilon_{k_h}) | k_i k_h \rangle_{AS}, \; k_i \leq k_M$$
$$= 0, \; k_i > k_M, \tag{11}$$

where the subscript $AS$ denotes antisymmetrized matrix elements. This prescription reduces the discontinuity in the single–particle spectrum as compared with the standard BHF choice $k_M = k_F$. The self–consistency scheme consists in choosing adequate initial values for the effective mass and $\delta$. The $G$–matrix obtained is then used to calculate the single–particle potential $u_i$, from which we obtain new values for $m^*$ and $\delta$. This procedure continues until these parameters vary little. The nucleon–nucleon potentials are defined by the meson–exchange potential models of the Bonn group, in table A.2 of ref. [5], versions A, B and C. By taking $k_C = k_M$ in equations (6) and (7) we can calculate single–particle energies and solve the gap equation in the same model space. Although BCS theory within the decoupling approximation does not dictate the choice of single–particle energies, this model-space approach is a consistent and very satisfying method where all quantities of interest are calculated in a systematic manner within a model space containing the physically relevant degrees of freedom. We should point out that in the energy gap calculations we have used the single–particle spectrum defined by eqs. (10) and (11), and not the effective mass approximation (9).

Having obtained single–particle energies, we proceed to solve eqs. (6) and (7). We choose $k_C = k_M$, and solve the gap equation in the model–space in which the single–particle energies have been calculated. The solution of equations (6) and (7) can be simplified by noting that for $k > k_F$, the dominating contribution to the quasiparticle energy $E_k$ comes from the term $(\epsilon_k - \epsilon_{k_F})^2$. This is clearly seen from fig. 3, where the quasiparticle energies are plotted for $k_F = 0.80$ fm$^{-1}$. Because $\Delta_k$ is of the order 1 MeV, while $E_k$ is of the order 100 MeV, we can safely neglect $\Delta_k$ for $k > k_M > k_F$. Thus equation (6) is decoupled from equation (7), and we can solve the linear equation for $\widetilde{V}$ by matrix inversion before proceeding to solve the gap equation. The gap equation in the continuum limit becomes

$$\Delta_k = -\frac{1}{\pi} \int_0^{k_M} dk' k'^2 \widetilde{V}_{kk'} \frac{\Delta_{k'}}{E_{k'}}. \tag{12}$$

Having obtained the effective interaction $\widetilde{V}$, we solve equation (12) by iteration. As shown in fig. 4 the integrand in the gap equation has a relatively sharp peak at the Fermi momentum, so in the numerical treatment we have used 50 Gauss–Legendre integration points in each of the intervals $[0, k_F + \eta]$ fm$^{-1}$ and $[k_F + \eta, k_M]$ fm$^{-1}$ in order to get numerically stable results, where $\eta = 0.1$ fm$^{-1}$ was found to be an adequate choice. Fig. 4 shows that it is necessary to use a rather large cutoff in the momentum space integration in the equation for $\widetilde{V}$.

To summarize: First we calculate renormalized single–particle energies using the MBHF method within a model space which size is determined by the parameter $k_M$. The next step is to evaluate the effective interaction $\widetilde{V}$, which sums up pairing excitations to states outside the model space. Finally, the gap equation is solved with this effective interaction as pairing interaction within the model space. This method is both numerically efficient and stable, and also allows us to construct an effective pairing interaction without double counting of two–particle correlations.

## III. PURE NEUTRON MATTER

As a first application we consider pure neutron matter at the relatively low densities found in the inner crusts of neutron stars. For simplicity we neglect the presence of finite nuclei in the crust, a presence which in refs. [18,19] was found to reduce considerably the neutron energy gaps compared with the pure neutron matter case. The calculation of energy gaps for realistic crust models is difficult, as one has to treat consistently the finite nuclei and the surrounding neutron gas. The main problem is to calculate pairing gaps in finite nuclei, and the validity of the local density approximations used so far has been questioned. In this section we therefore choose to focus on the more transparent case of pure neutron matter.

### A. Choice of model–space

An important question is the choice of the parameter $k_M$ which determines the size of the model space. Previous investigations for the MBHF-method [20] have shown that the results are stable in a region around $k_M = 3.0$ fm$^{-1}$. Baldo et al. [4] found their results to be stable with respect to variations of $k_M$, however, they did not give results for the $k_M$–dependence of the energy gap $\Delta_F \equiv \Delta_{k=k_F}$. In fig. 5 we have plotted the energy gap at the Fermi surface, $\Delta_F$, calculated with the Bonn A potential, as a function of $k_M$ for various choices of $k_F$. The figure shows that the energy gap is virtually independent of $k_M$ for $k_M$ in the interval $1.0 \text{ fm}^{-1} < k_M < 4.0 \text{ fm}^{-1}$. As $k_M \to k_F$ we get trouble due to our approximation $\Delta_{k>k_M} = 0$ causing the energy denominator in equation (6) to almost vanish at $k_F$. However, this problem is avoided in a fully self-consistent approach to equations (6) and (7), where $\Delta_F$ gives a nonzero contribution to the energy denominator at $k_F$. For $k_M \approx 4.0 \text{ fm}^{-1}$ large contributions from the repulsive core of $V$ are included directly in the gap equation, and ultimately as $k_M \to \infty$, the effective interaction becomes equal to the bare interaction. We notice that the numerical instabilities grow with increasing density. This is due to the increasing importance of the repulsive core. The results in this paper have been obtained with $k_M = 3.0 \text{ fm}^{-1}$. Figures 6 and 7 show the $k_M$-dependence of the energy gap for three selected values of $k_F$ for Bonn B and Bonn C. In these calculations we included values of $k_M$ close to $k_F$. It is seen that as $k_M$ approaches $k_F$ the value of the energy gap decreases rapidly, due to the discontinuity in the single-particle spectrum. We see that the choice $k_M = 3.0 \text{ fm}^{-1}$ is right in the center of the stable region.

The choice of upper limit for the numerical integration in equation (6) has also been shown to be important, and should according to ref. [17] be chosen greater than 50–60 fm$^{-1}$ for the Paris potential, due to the complicated off-diagonal behaviour of this potential in momentum space. Accordingly, we have made sure that all results in this paper are stable with respect to the choice of upper limit for the integration in eq. (6).

### B. Energy gaps in pure neutron matter

We have solved the gap equation in pure neutron matter with the potential models A, B and C of table A.2 of ref. [5]. The results in fig. 8 show that the gap at the Fermi surface, $\Delta_F \equiv \Delta_{k_F}$, is more or less the same for the three potential models. There is a slight difference in the maximum value of the energy gap, which is $\Delta_{k_F=0.80} = 2.71$ MeV for Bonn A, $\Delta_{k_F=0.81} = 2.76$ MeV for Bonn B and $\Delta_{k_F=0.81} = 2.75$ MeV for Bonn C. The energy gap disappears for $k_F = 1.43 \text{ fm}^{-1}$ with all potential models. These results should be compared with those of Baldo et al. [4], where a separable form of the Paris interaction gave the results $\Delta_{k_F=0.85} = 2.76$ MeV, $\Delta_{k_F \approx 1.4} = 0$, and similar results were found with the Argonne $V_{14}$ interaction. Our results agree with these results, as was to be expected, since all realistic nucleon-nucleon interactions reproduce the $^1S_0$ scattering data. Moreover, in this channel, only central force components contribute to the NN potential, so that more complicated off-shell effects from tensor forces are non-present. Thus at the two-body level all realistic NN-interactions should give similar results for the $^1S_0$ energy gap in pure neutron matter. The differences, apart from different choices for the single particle energies, may be traced to the off-shell behaviour of the interactions, which has been shown in ref. [4] to influence the value of the energy gap at $k_F$. The advantage of the approach used here over that in [4] lies in the choice of MBHF instead of BHF single-particle energies, by which we avoid a discontinuity in the single-particle spectrum in the space where the gap equation (7) is solved.

We are also interested in the gap function $\Delta_k$. To obtain the gap function for values of $k$ above $k_M$ we need to solve equations (6) and (7) self-consistently. This will also provide us with a check of the accuracy of the approximation $\Delta_k = 0$, $k > k_M$. The self-consistent solution of equations (6) and (7) is obtained by using the solution for $\Delta_k = 0$, $k > k_M$ to calculate an initial approximation for $\Delta_{k>k_M}$ from the gap equation (7). With this initial approximation we calculate a new effective interaction from equation (6) and then obtain a new approximation to $\Delta_k$ from equation (7). The procedure is continued until self-consistency is achieved. The results for the energy gap $\Delta_F$ were found to change insignificantly from the results with $\Delta_{k>k_M} = 0$, as seen from table 1. In figure 9 the gap function is shown for $k_F = 0.81 \text{ fm}^{-1}$, calculated with the Bonn A potential. The gap function is seen to vary rapidly around $k_F$ and is negative over a wide range of momenta. However, contrary to the results with the Paris interaction in ref. [4], we find that the gap function decays rather quickly and shows little oscillation at large $k$. The difference is most likely due to the different off-diagonal momentum dependencies of the potentials.

### C. Correlated wave functions

Another interesting quantity is the BCS pair wave function. For a given Fermi momentum $k_F$, this function can be calculated from the gap function $\Delta_k$ using the expression

$$\Psi^{BCS}(r) = \int_0^\infty dk\, k^2 j_0(kr) \frac{\Delta_k}{\sqrt{(\epsilon_k - \epsilon_{k_F})^2 + \Delta_k^2}}, \tag{13}$$

which is the Fourier transform of the expectation value $\langle c | a_{\mathbf{k}\uparrow}^\dagger a_{-\mathbf{k}\downarrow}^\dagger | c \rangle$, with $|c\rangle$ denoting the usual BCS ground state, and $a_{\mathbf{k}\uparrow}^\dagger$ creating a (normal) particle with momentum $\mathbf{k}$ and spin projection $m_S = +\frac{1}{2}$ [4] The normalization

constant has been omitted. Fig. 10 shows this wave function together with the Brueckner two–particle wave function in the $^1S_0$ channel for two particles with opposite momenta at the Fermi surface, given by

$$\Psi_0(r, k_F) = j_0(k_F r) \qquad (14)$$
$$+ \frac{2}{\pi} \int_{k_F}^{\infty} dq\, q^2 j_0(qr) \frac{Q(q, K=0)}{\omega - H_0(q, 0)} \langle q | G_0(K=0, \omega) | k_F \rangle,$$

where $\omega = \frac{1}{m^*} k_F^2 + 2\delta$ is the starting energy, $Q(q, 0)$ is the Pauli operator, $\mathbf{K}$ is the center of mass momentum, and $G_0$ is the $^1S_0$ component of the $G$–matrix. Eq. (14) is a special case of the more general expression found in ref. [6]. We see that the behaviour of the two functions is very similar for small $r$, and this supports the view that the gap equation plays a part in accounting not only for the long range pairing correlations, but also for the short range two–particle correlations. A similar result was found by Baldo et al. [17] for pairing in the $^3S_1$ channel in symmetric nuclear matter.

### D. Finite temperature

The gap equation at finite temperature $T$ is

$$\Delta_k(T) = -\frac{1}{\pi} \int_0^{k_M} dk'\, k'^2 \widetilde{V}_{kk'} \frac{\Delta_{k'}(T)}{E_{k'}} \tanh\left(\frac{E_{k'}}{2 k_B T}\right) \qquad (15)$$

where $k_B$ is Boltzmann's constant. To solve equation (15) we use the same approach as for $T = 0$. We make the simplification of ignoring the temperature dependence in $\widetilde{V}$, because in the temperature range of interest $k_B T \approx 1$ MeV, while for $k > k_M$, $E_k$ is at least of order 100 MeV, and thus we can ignore the role of thermal excitations to states above $k_M$. Furthermore, following common practice, we use "frozen" single–particle energies, that is, we calculate them within the MBHF approach at $T = 0$. According to refs. [21,22], this should be a reasonable approximation. We therefore calculate an effective interaction from equation (6) and use this to solve the gap equation (15) by iteration, just as for $T = 0$.

In fig. 11 we show the temperature dependence of the energy gap at the Fermi surface, $\Delta_F(T)$, calculated with the Bonn A potential. Also shown in fig. 11 are the critical temperatures estimated from the popular weak–coupling approximation (WCA) [23]

$$k_B T_C \approx 0.57 \Delta_F(T=0) \qquad (16)$$

where $T_C$ is the critical temperature. Our results indicate that this approximation is reasonably good, but that it systematically underestimates the critical temperature. The WCA seems to decrease in quality as the density increases, although not as drastically as the WCA for the energy gap and the condensation energy, which in ref. [4] were found to deviate considerably from the self–consistent results. That the WCA for the critical temperature turns out to be fairly accurate can be understood by considering its derivation, as found in e.g. ref. [23]. The main assumptions are that $k_B T_C \ll \epsilon_F$ and that it is sufficient to integrate over a small band of states near the Fermi surface. For neutron matter, the first assumption is reasonable, the last assumption is, however, dubious, as can be seen from fig. 4. However, $\Delta_F$ in eq. (16) is really the WCA for the energy gap,

$$\Delta_F \approx 2\epsilon_F \exp\left(4\pi^2 \frac{\epsilon_F}{\widetilde{V}_{k_F k_F} k_F^3}\right) \qquad (17)$$

and if, as has been done here, it is replaced by the energy gap found by solving eq. (7), much of the error done in neglecting the integration over large momenta is compensated for.

### IV. PROTON PAIRING AT $\beta$ EQUILIBRIUM

We have also solved the gap equation for protons in neutron star matter at beta equilibrium, with and without muons. The proton fraction is determined by imposing the relevant equilibrium conditions on the processes

$$e^- + p \to n + \nu_e \qquad (18)$$

$$e^- \to \mu^- + \overline{\nu}_\mu + \nu_e \tag{19}$$

The conditions for $\beta$ equilibrium require that

$$\mu_n = \mu_p + \mu_e \tag{20}$$

where $\mu_i$ is the chemical potential of particle species $i$, and that charge is conserved

$$n_p = n_e \tag{21}$$

where $n_i$ is the particle number density for particle species $i$. If muons are present, the condition for charge conservation becomes

$$n_p = n_e + n_\mu \tag{22}$$

and chemical equilibrium in the process (19) gives the additional constraint

$$\mu_e = \mu_\mu \tag{23}$$

Throughout we have assumed that neutrinos escape freely from the neutron star. The proton and neutron chemical potentials are determined from the energy per baryon, calculated self–consistently in the MBHF approach. The electron chemical potential, and thereby the muon chemical potential, is then simply given by $\mu_e = \mu_n - \mu_p$. The Fermi momentum of lepton type $l = e, \mu$ is found from

$$k_{F_l}^2 = \mu_l^2 - m_l^2$$

and we get the particle density using $n_l = k_{F_l}^2/3\pi^2$. The proton fraction is then determined by the charge neutrality condition (22). Having calculated proton fractions and neutron and proton single–particle energies, we can solve the gap–equation for protons in the same manner as we did for neutrons in the previous section. In constructing the effective pairing interaction $\widetilde{V}$ for protons we neglect neutron–proton interactions, since the Fermi surfaces of neutrons and protons are widely separated. All results in this section have been obtained using the Bonn A potential. In fig. 12 the proton fractions obtained with and without muons are shown. The proton fraction is given as $n_p/n_B$, where $n_B = n_n + n_p$ is the total baryonic number density. The muons enter the equation of state at a baryon density $n_B \approx 0.15$ fm$^{-3}$, and as a result the proton fraction increases. We also show the results of MBHF calculations of the proton effective mass in fig. 13. The faster decrease in the proton effective mass when muons are included is due to the larger proton density obtained with muons. Figure 14 shows the proton energy gap as function of the total baryonic density and in fig. 15 as a function of the proton Fermi momentum. In fig. 15 we also show the neutron energy gap calculated at $\beta$ equilibrium. We see from these results that the inclusion of muons leads to a proton energy gap that decreases faster, but has a larger maximum value compared to the case with no muons. The faster decline of the energy gap is mainly due to the lower proton effective mass obtained with muons included, as shown in figure 13. The slightly larger maximum value of the energy gap is probably due to the increase in the proton density when muons are included, i.e. a proton "sees" more protons with which it can pair, and at the density where the proton gap reaches its maximum, this effect is greater than the effect of the increased density on the proton effective mass. The neutron energy gap is only slightly modified, due to the very low proton fraction at the baryon densities where neutron $^1S_0$ pairing is possible. We should mention that the presence of muons has little influence on the energy per particle and the resulting maximum masses for neutron stars, as discussed in [24].
To the best of our knowledge, there exists no previous calculations of the proton energy gap with muons included. For the results without muons, the most natural work to compare our results with is that of Baldo *et al.* [25]. There the Argonne $V_{14}$ interaction [26] was used, and proton effective masses calculated in the BHF approximation. They found a maximum proton energy gap of approximately 0.8 MeV, and that $\Delta_F(k_F^p) = 0$ at $k_F^p \approx 0.9$ fm$^{-1}$. The differences between these values and those found in this work are probably due to the smaller proton effective mass ratios and smaller proton fractions obtained in ref. [25]. Amundsen and Østgaard [10] obtained a maximum energy gap of 0.3 MeV using proton effective masses taken from Takatsuka [27], where a maximum proton gap of 0.5 MeV for $k_F^p = 0.5$ fm$^{-1}$ and a closing density of $k_F^p = 0.8$ fm$^{-1}$ was found. These low values are probably due to proton effective mass ratios in the range 0.5–0.7 being used in these two works. Chen *et al.* [15] obtained a maximum value of about 1.1 MeV within a variational (CBF) approach. However, as pointed out in that work, polarization effects will probably have a profound influence on the magnitude of the proton energy gap. Recent studies within the polarization–potential approach [28] indicate that medium effects can reduce the size of the energy gap by as much as a factor of three.

## V. DISCUSSION

The energy gaps calculated in this work are lower than those obtained with variational [10,15] and $G$-matrix [12] approaches. However, the results are not directly comparable, not only due to different choices of NN potentials, single-particle spectra, different calculations of proton fractions etc., but most importantly because they are obtained within different theoretical frameworks. As far as approaches using the Brueckner $G$-matrix is concerned, it is hard to see how these procedures can be justified in a strict manner. In perturbation theory, the $G$-matrix emerges as a natural reformulation of the original divergent perturbation series, but no similar reformulation of the pairing problem has been attempted within standard BCS theory. The CBF method used in refs. [11,15] employs basis states designed to handle short-range correlations, and at the pure variational level, the energy gaps obtained in this framework are larger than the bare interaction BCS results by some 0.5 MeV. It does not seem to be clear that double counting of two-particle correlations is avoided within the pure variational CBF approach, that is, that there is a clear division between the role of the pair correlation function and the correlation factors introduced in the CBF method. If medium polarization effects are omitted, Green's functions methods strongly indicate [29] that the pairing matrix elements of the bare nucleon-nucleon interaction should be used in the gap equation. The reformulation of the gap equation through the coupled equations (6) and (7) together with the wave functions in fig. 10 provide evidence for the role played by the gap equation in taking care of the short-range correlations. The advantage of the model-space approach used in this work is that we can explicitly show that double counting is avoided. Furthermore, the gap equation is solved in a space where we use renormalized single-particle energies; free particle energies are only used in the construction of the effective interaction $\widetilde{V}$. The renormalized interaction $\widetilde{V}$ makes the numerical treatment of the gap equation considerably easier, because the highly nonlinear part of the equation is separated from the almost linear part. We stress the need for extending the momentum space integration in the gap equation to large values of $k$. Thus the weak-coupling approximation used in ref. [13] is probably not appropriate, as also remarked in refs. [4,15], and tends to underestimate the energy gap. However, as they use a $G$-matrix or equivalent effective interaction in the construction of their quasiparticle interaction, this may compensate for the error caused by the weak-coupling approximation. The status of the results found in ref. [13] seems therefore to be uncertain.

The inclusion of muons has a small, but significant effect on the proton energy gap. However, the proton energy gaps must be considered tentative, as we have neglected polarization effects. Furthermore, relativistic effects become important at the densities where proton pairing is possible. We hope to include both these effects in a future investigation. The proton fraction is of significant importance, and this again depends sensitively on the symmetry energy, which is a poorly known quantity. Albeit all these reservations, all calculations, including the one in this work, have so far been consistent with the existence of superconducting protons in neutron stars.

Since all calculations, including the ones in this work, give neutron energy gaps of order 1 MeV, one can conclude that neutrons in the inner crusts of neutron stars are likely to form a superfluid. Corrections to the simple and naive model of a uniform neutron gas interacting via two-body forces are probably not large enough to destroy neutron superfluidity. However, as stated in the beginning of this paper, more realistic calculations are needed in order to study dynamical and thermodynamical effects in the inner crusts. The recent finding that the lattice nuclei in this region may take on non-spherical shapes [30,31] further complicates the picture. The impact of this fact on the specific heat of crust matter and the vortex pinning energy has been found to be significant [32]. The last quantity is of interest in models for pulsar glitches.

The proton energy gaps obtained in this work have recently been used in a calculation of rates for the modified URCA processes [33]. Not surprisingly, the rates were found to be significantly suppressed by proton superconductivity. The analysis of Page [34] indicates that it is hard to distinguish fast from slow cooling scenarios; in both cases agreement with the observed temperature of e.g. Geminga is obtained if baryon pairing is present in most, if not all of the core of the star.

## VI. CONCLUSION

The reformulation of the BCS gap equation advocated by Baldo *et al.* [4,17] allows for a consistent treatment of pairing correlations without double counting. One also gets the advantage of using a renormalized interaction, so that the main effects of the repulsive core of the NN interaction can be integrated out before proceeding to solve the gap equation. The additional consistency in solving the gap equation in the same model space in which the single-particle energies are calculated is also an advantage of this method. However, medium polarization effects on the effective pairing interaction must be taken into account, and work in this direction is in progress. From our point of view, the model-space approach to pairing correlations seems to offer possibilities also for a consistent treatment of medium polarization through the inclusion of particle–hole ring diagrams.
This work has been supported by the Research Council of Norway (NFR) under the Programme for Supercomputing. One of us, (MHJ), thanks the NFR and the Istituto Trentino di Cultura, Italy, for financial support.

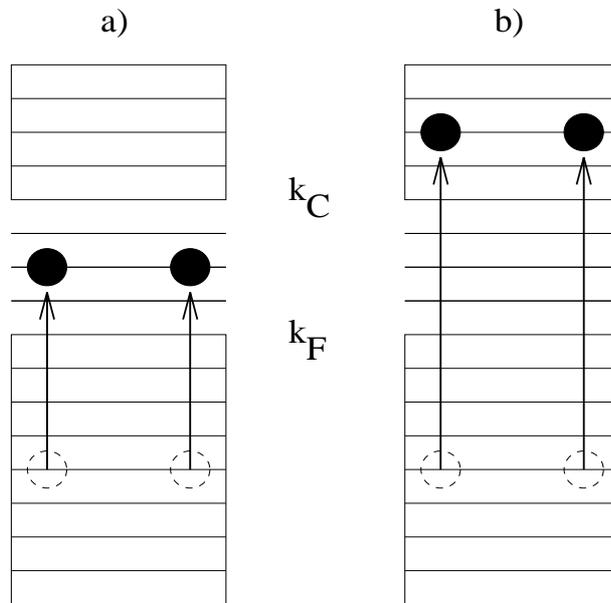

FIG. 1. Two–particle excitations involved in pairing. Those in a) are included in the gap equation (6), and those in b) are included in the effective interaction (5).

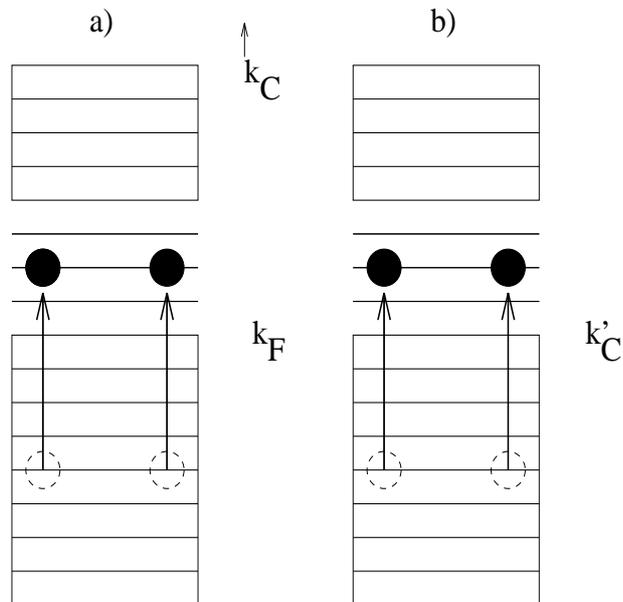

FIG. 2. If we take $k'_C = k_F$ in the effective interaction, while $k_C = \infty$ in the gap equation, we get the situation shown above, where the two–particle excitation in a) and b) is included in both equation (5) and in equation (6).

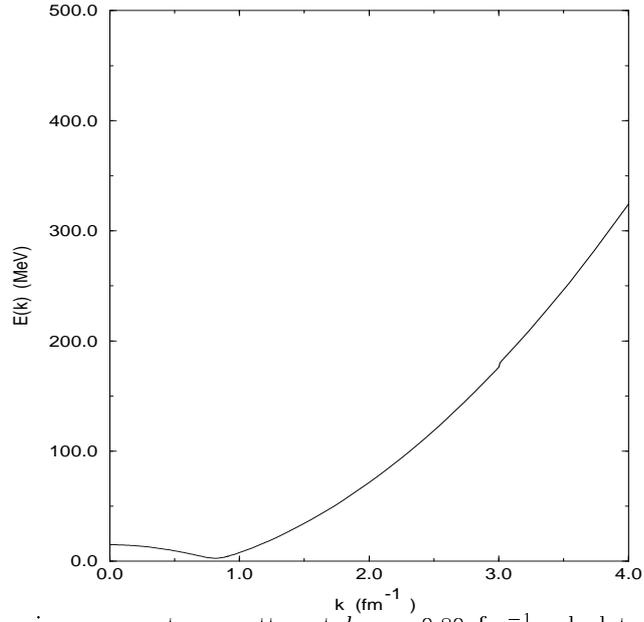

FIG. 3. Quasiparticle energies in pure neutron matter at $k_F = 0.80$ fm$^{-1}$ calculated with the Bonn A potential and $k_M = 3.0$ fm$^{-1}$.

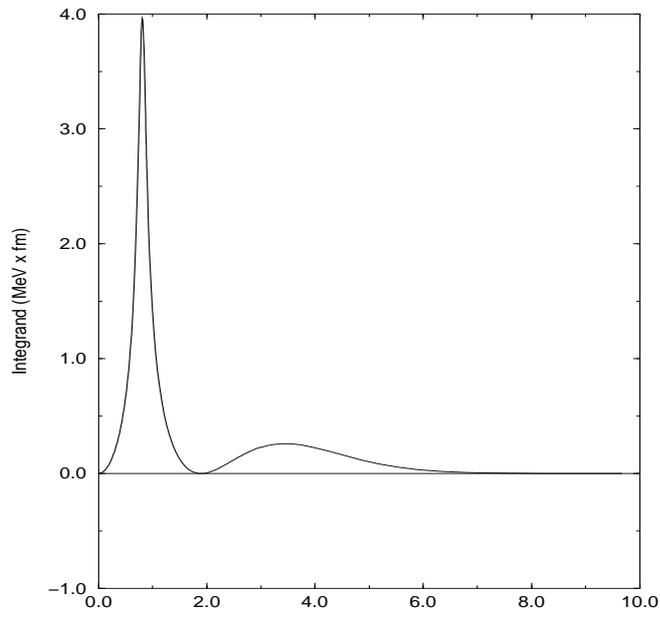

FIG. 4. The integrand in the gap equation for $k_F = 0.80$ fm$^{-1}$ and $k = 0.90$ fm$^{-1}$, calculated with the Bonn A potential.

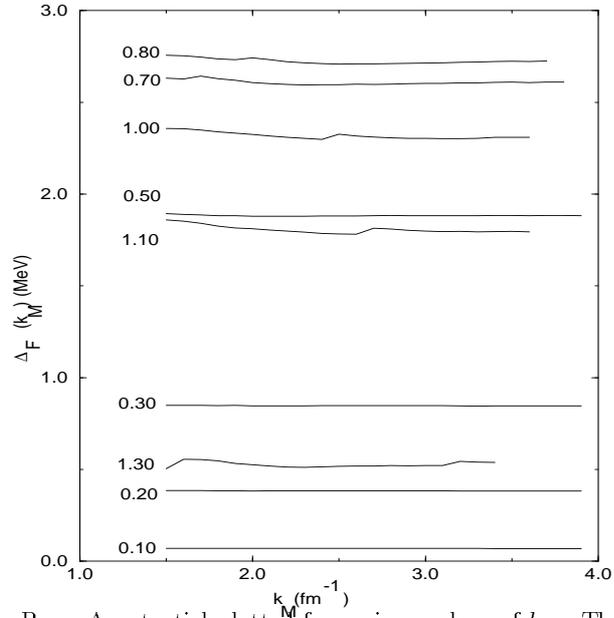

FIG. 5. Energy gap $\Delta_F$ with the Bonn A potential plotted for various values of $k_M$. The numbers next to the lines are the corresponding values of $k_F$.

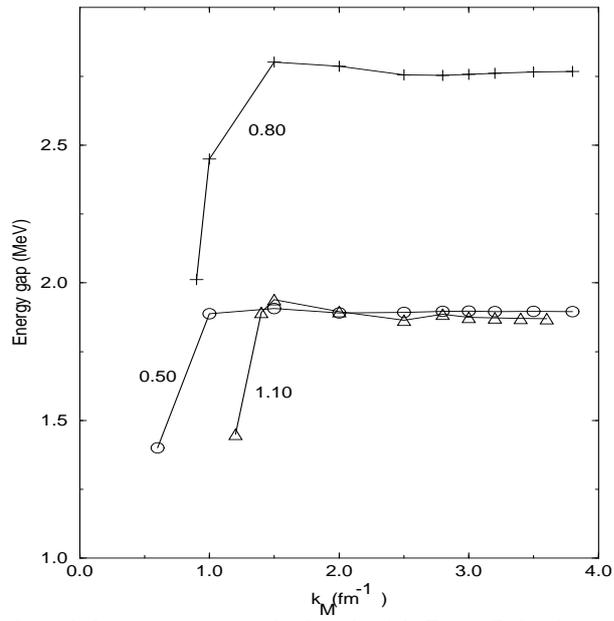

FIG. 6. Dependence upon $k_M$ of the energy gap calculated with Bonn B for $k_F = 0.50$, $0.80$ and $1.10$ fm$^{-1}$.

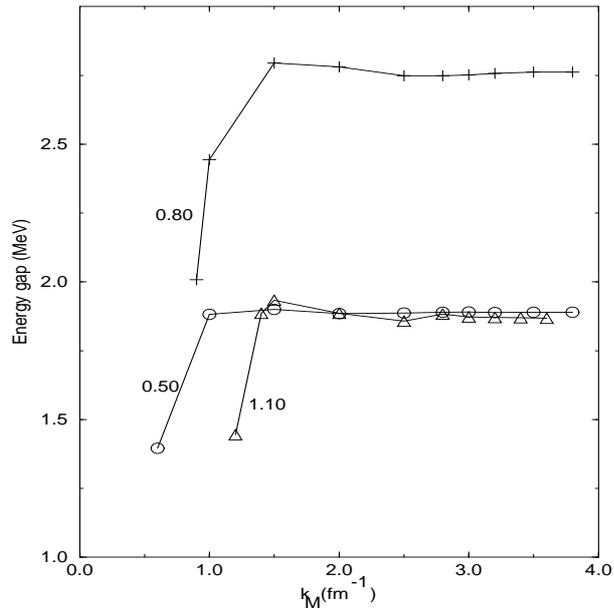

FIG. 7. Dependence upon $k_M$ of the energy gap calculated with Bonn C for $k_F = 0.50$, 0.80 and 1.10 fm$^{-1}$.

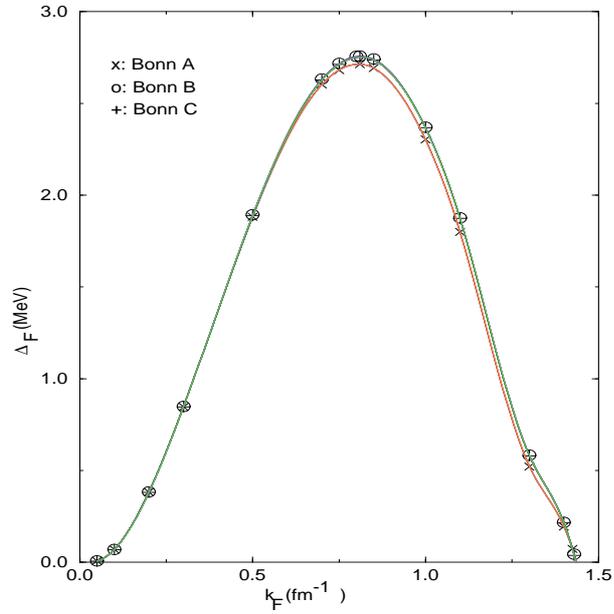

FIG. 8. The energy gap $\Delta_F$ calculated with the Bonn A, B and C potentials.

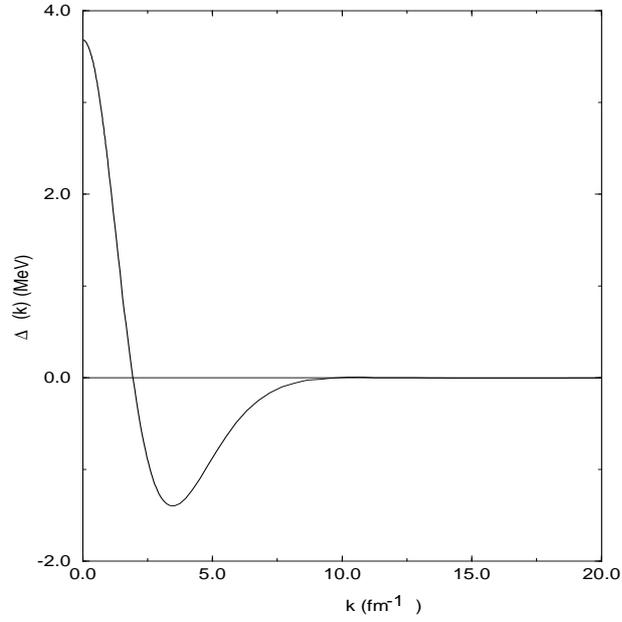

FIG. 9. Gap function $\Delta(k) \equiv \Delta_k$ calculated with the Bonn A potential for $k_F = 0.81$ fm$^{-1}$.

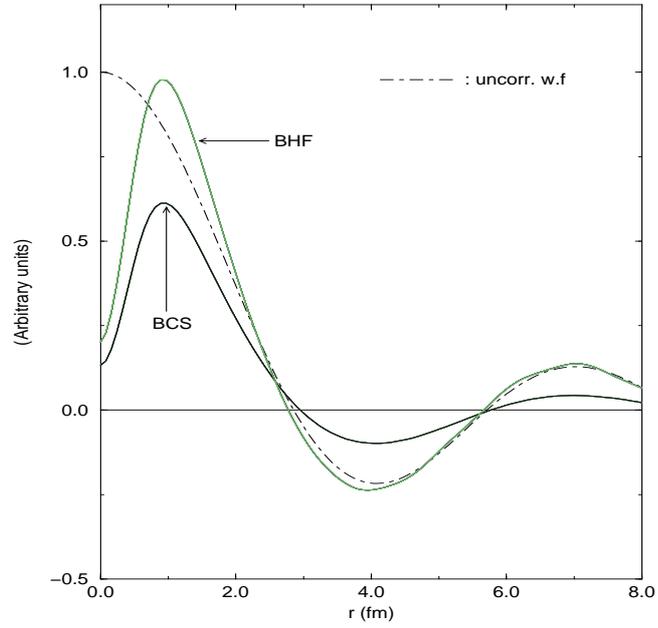

FIG. 10. The wave functions $\Psi^{BCS}$ and $\Psi_0$ for $k_F = 1.1$ fm$^{-1}$.

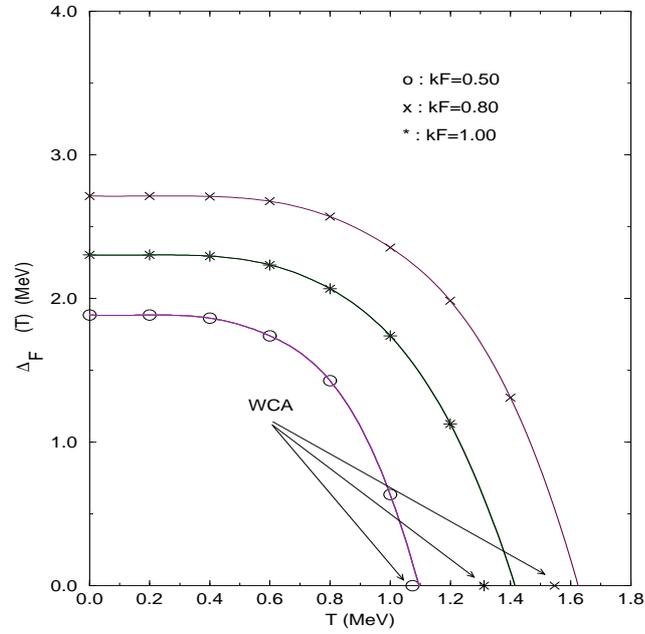

FIG. 11. Temperature dependence of $\Delta_F$ for the values of $k_F$ (in fm$^{-1}$) indicated in the figure. The corresponding weak–coupling estimates for the critical temperatures are also indicated.

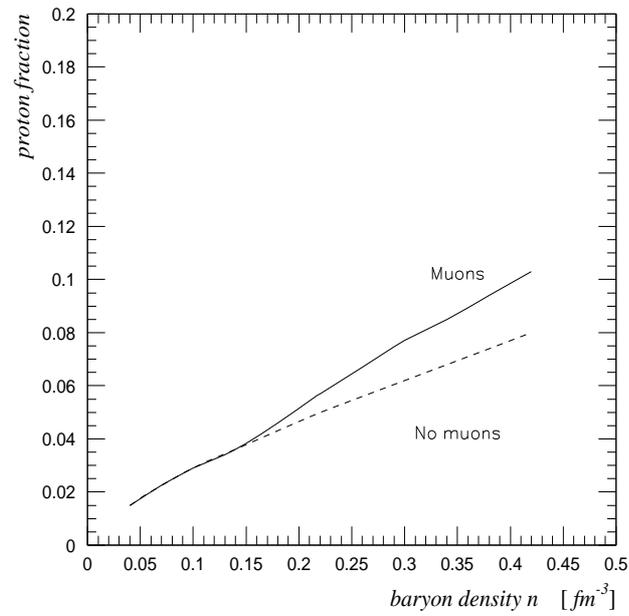

FIG. 12. Proton fractions in neutron star matter at $\beta$ equilibrium.

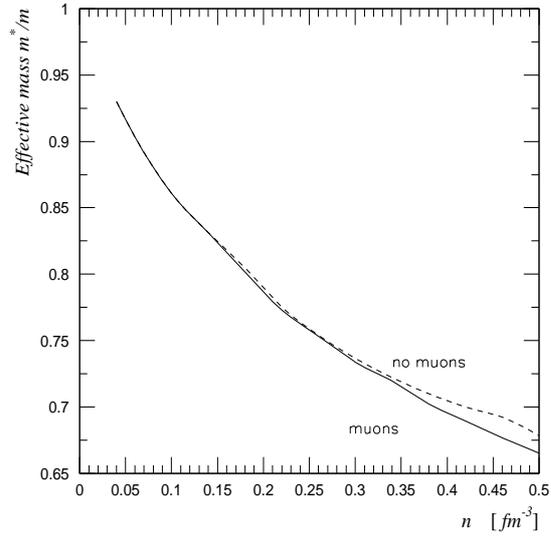

FIG. 13. Proton effective mass ratio $m^*/m$ in neutron star matter at $\beta$ equilibrium, shown as a function of baryon density.

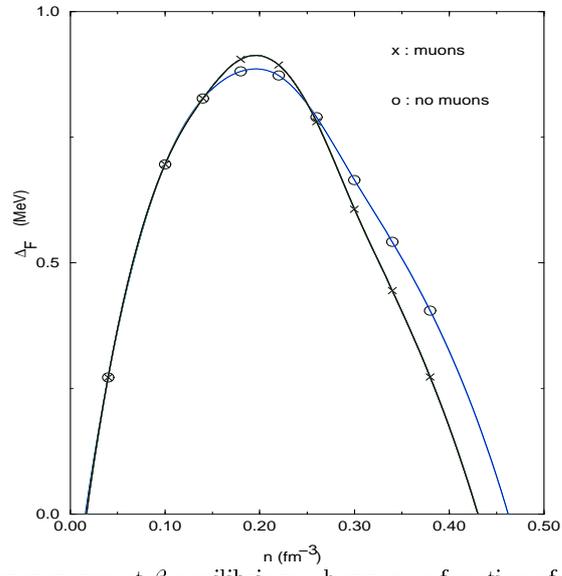

FIG. 14. Proton energy gap at $\beta$ equilibrium, shown as a function of baryon density.

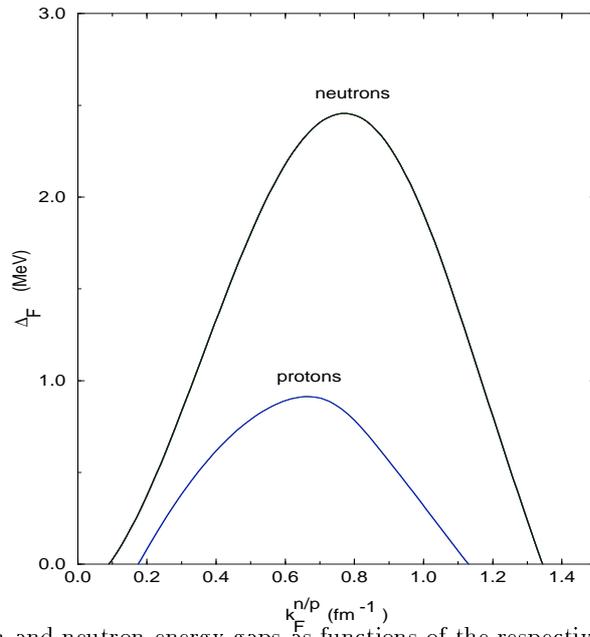
FIG. 15. Proton and neutron energy gaps as functions of the respective Fermi momenta.

| | |
|---|---|
| | 0.05 |
| | 0.008 |
| | 0.008 |
| | 0.10 |
| | 0.070 |
| | 0.070 |
| | 0.20 |
| | 0.383 |
| | 0.384 |
| | 0.30 |
| | 0.847 |
| | 0.847 |
| | 0.50 |
| | 1.883 |
| | 1.882 |
| | 0.70 |
| | 2.603 |
| | 2.600 |
| | 0.81 |
| | 2.714 |
| | 2.711 |
| | 0.85 |
| | 2.693 |
| | 2.689 |
| | 1.00 |
| | 2.304 |
| | 2.301 |
| | 1.10 |
| | 1.800 |
| | 1.798 |
| | 1.30 |
| | 0.522 |
| | 0.522 |

TABLE I. Energy gaps calculated with Bonn A. Here $\Delta_F^{(1)}$ has been obtained with the approximation $\Delta_{k>k_M} = 0$, while $\Delta_F^{(2)}$ is the complete self–consistent solution.

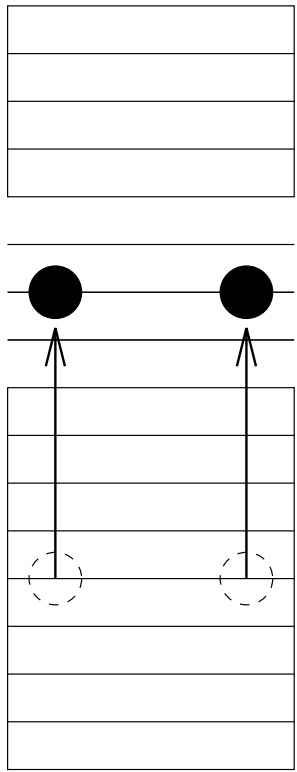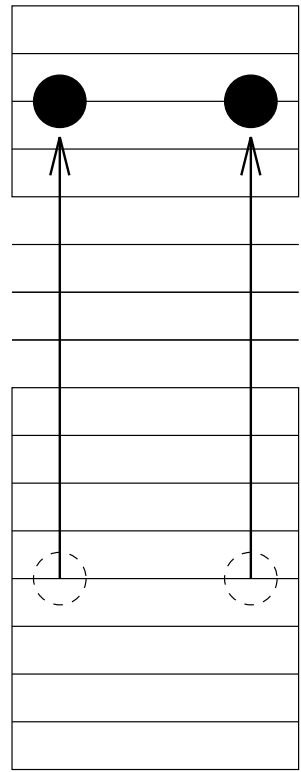

$k_C$

$k_F$

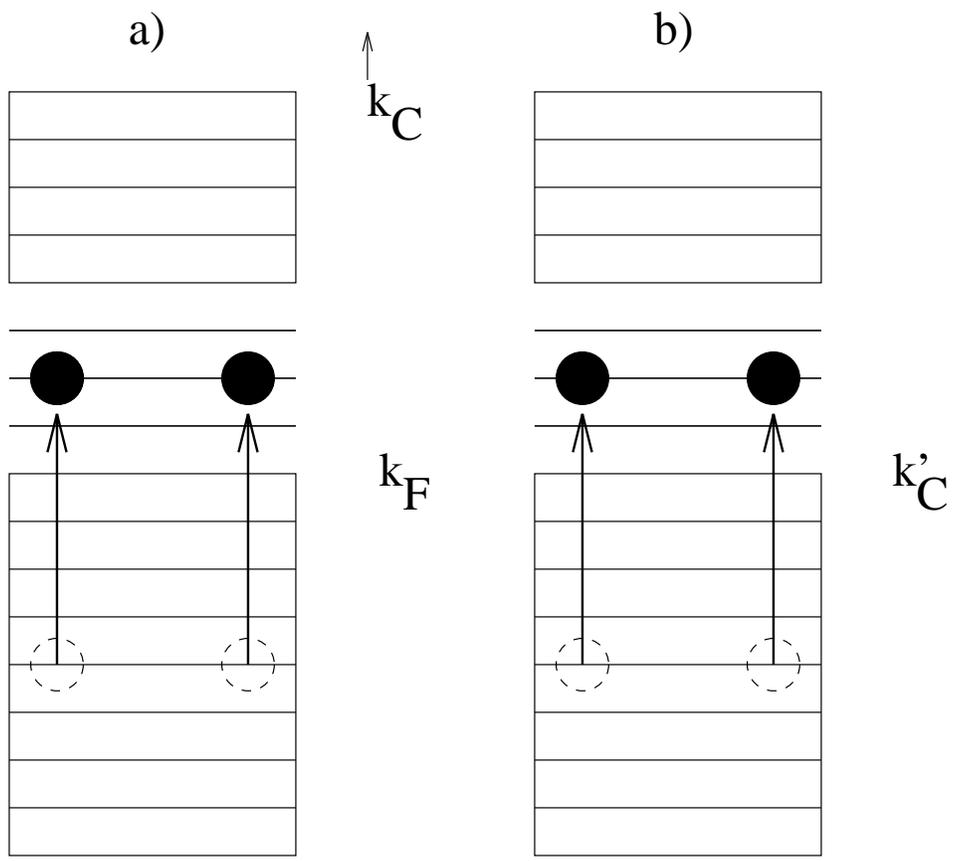

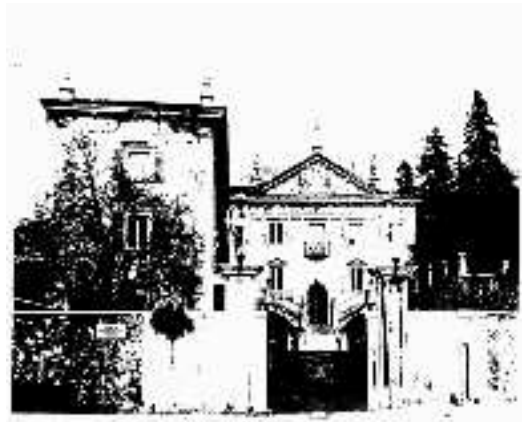

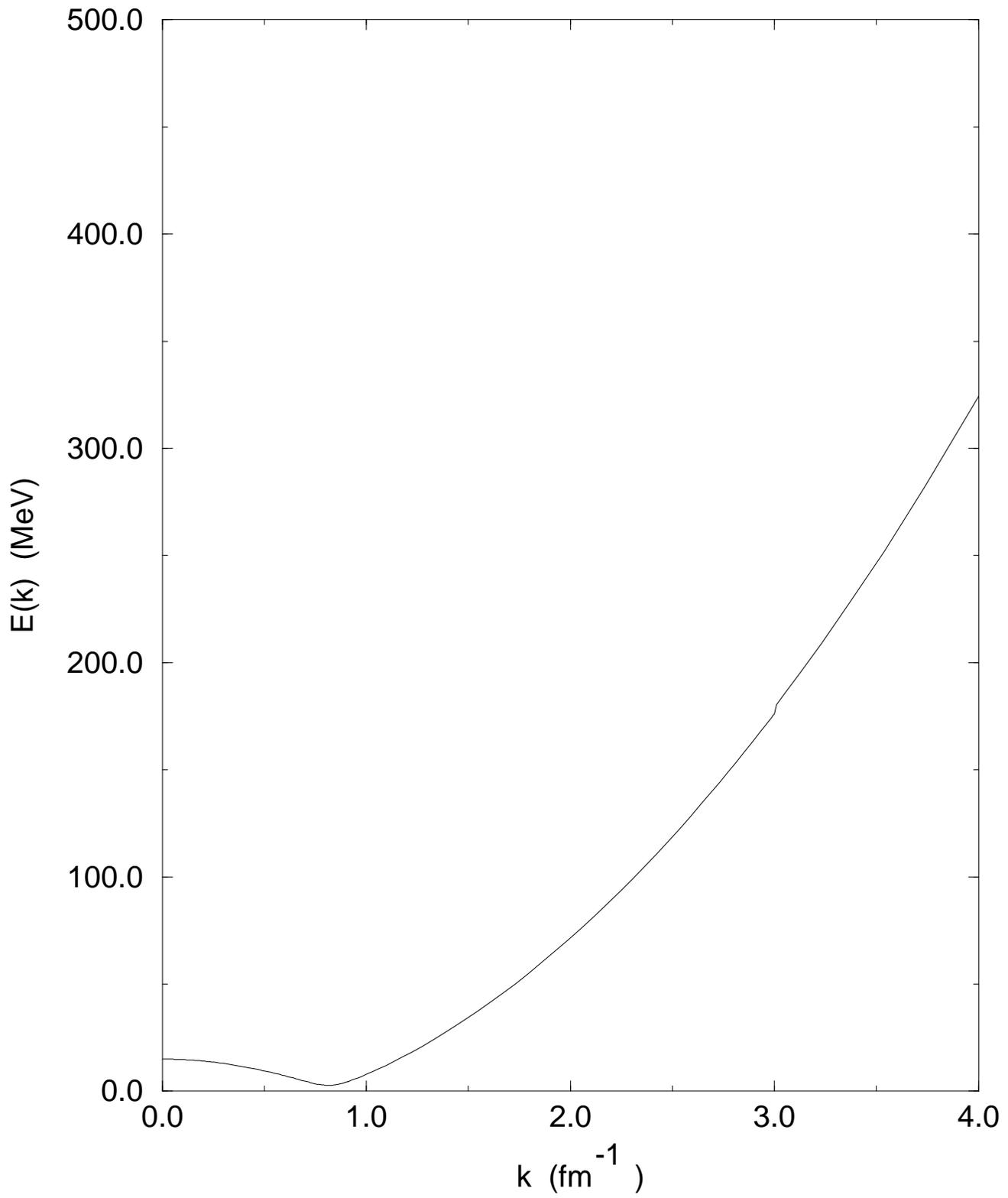

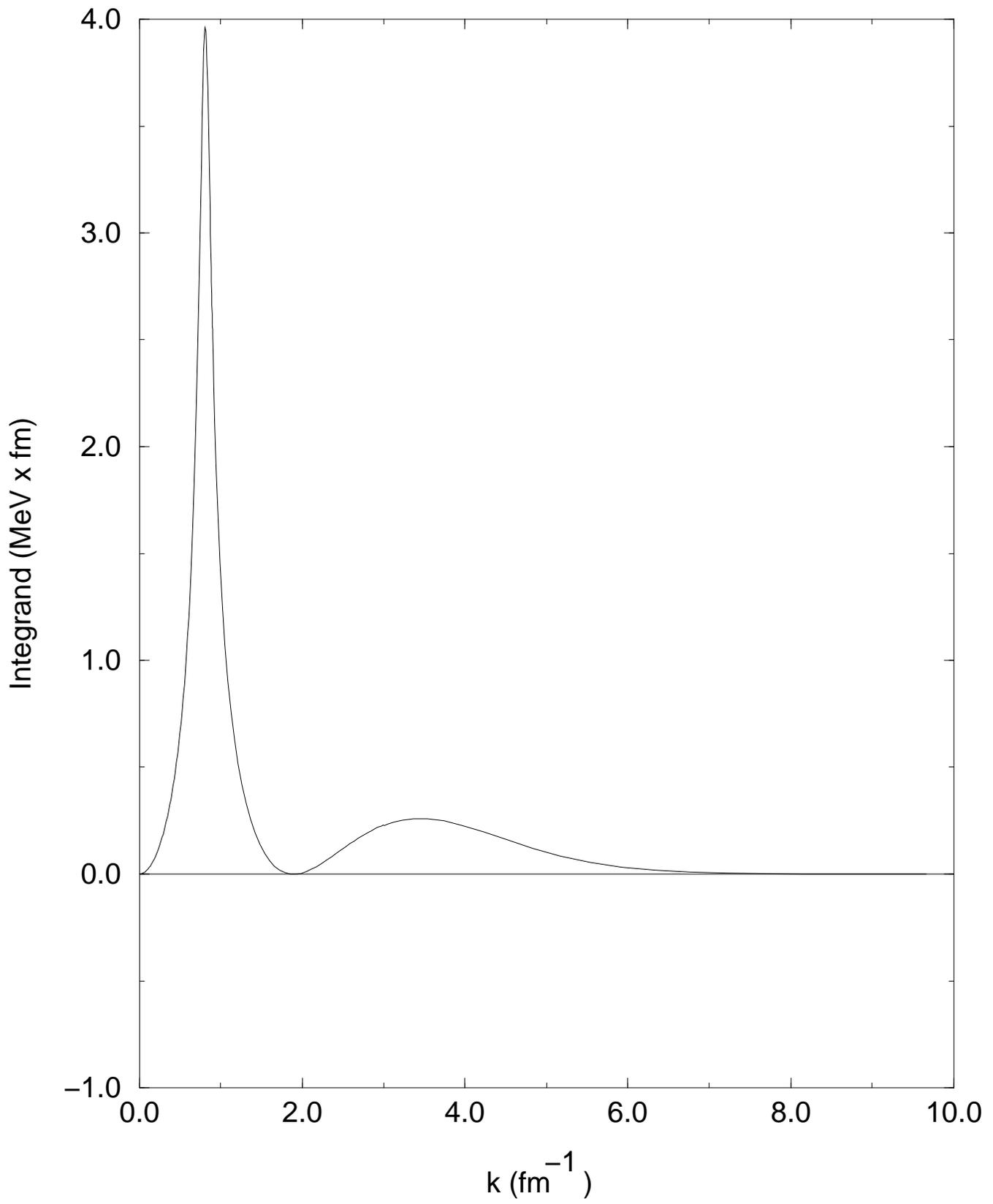

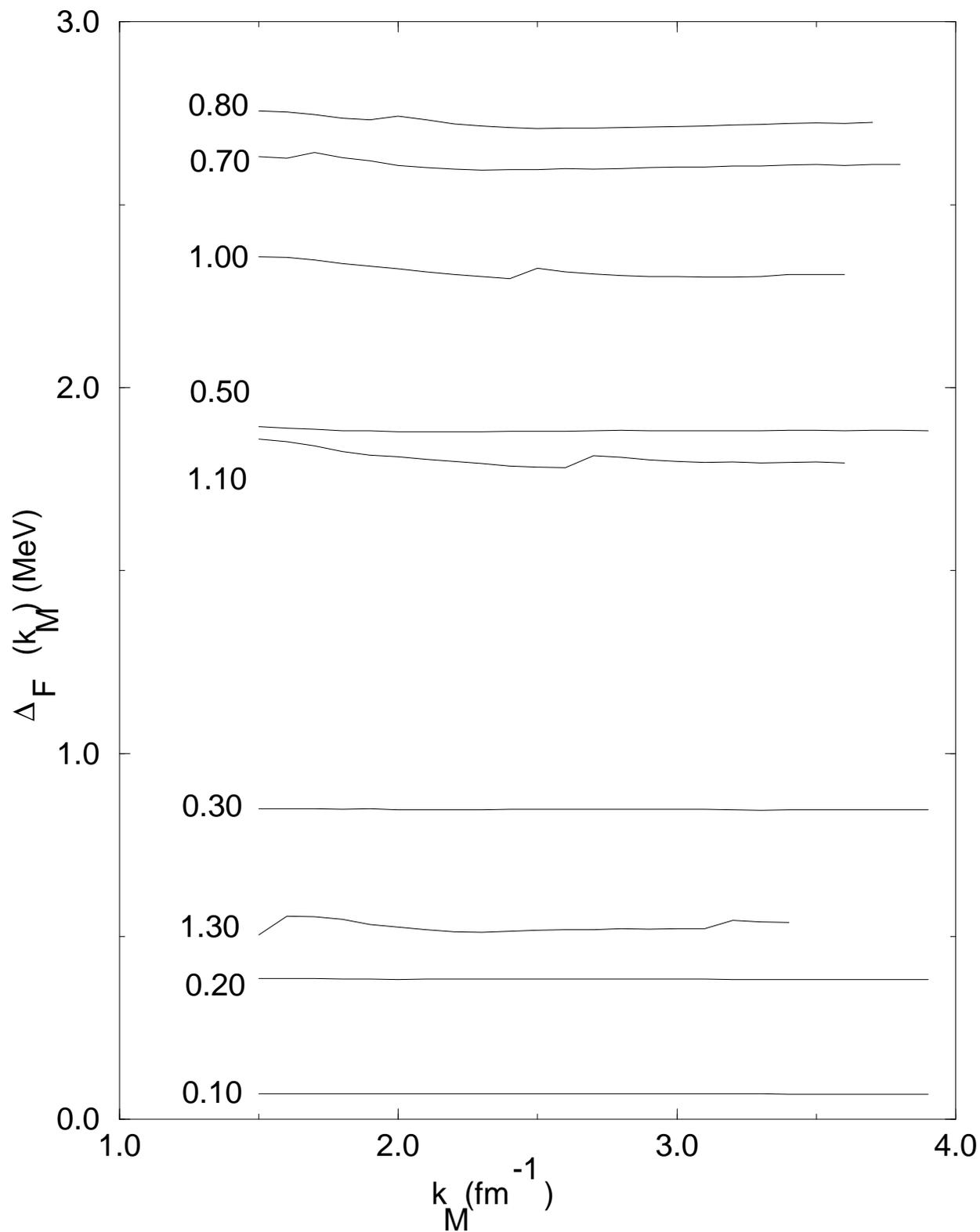

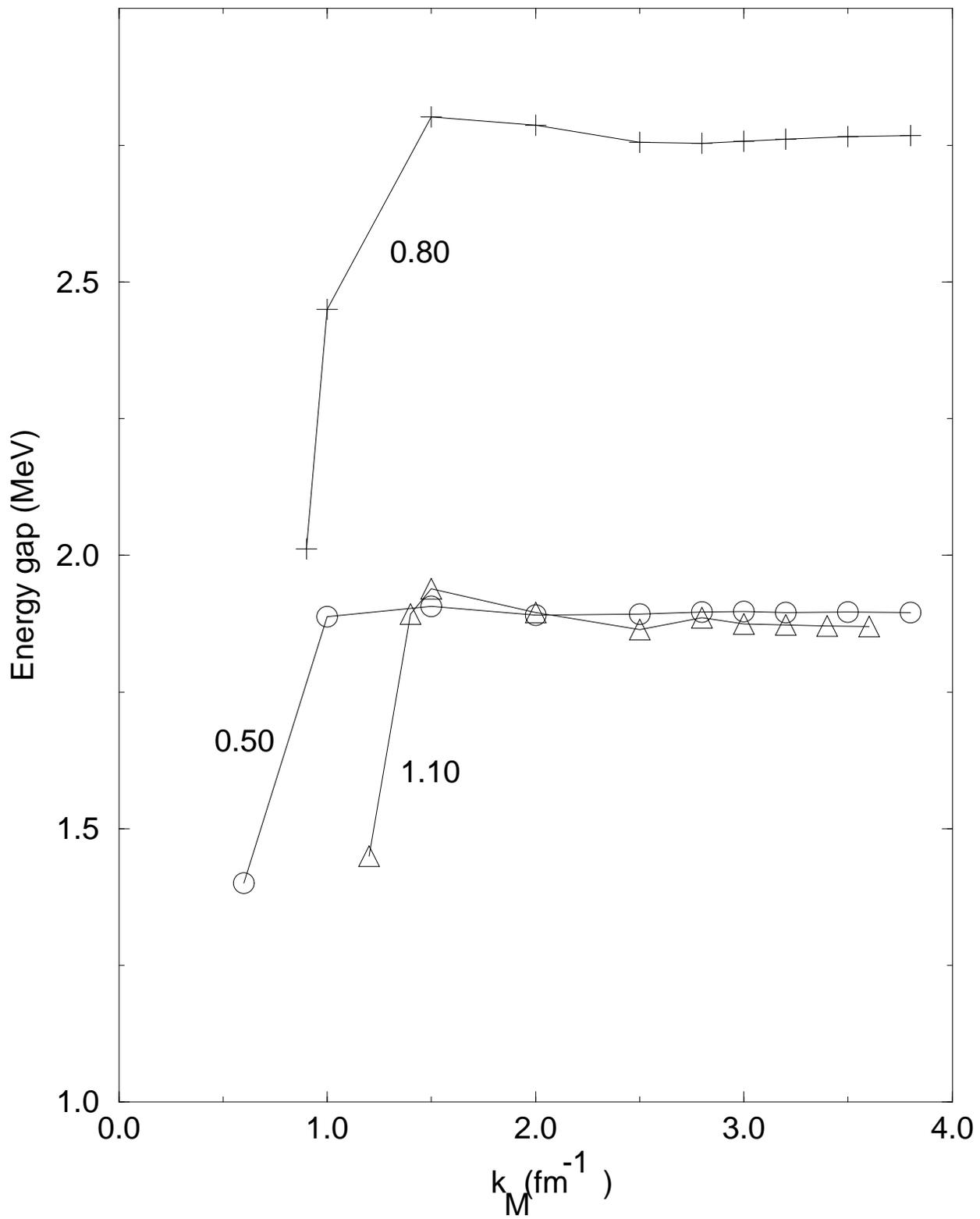

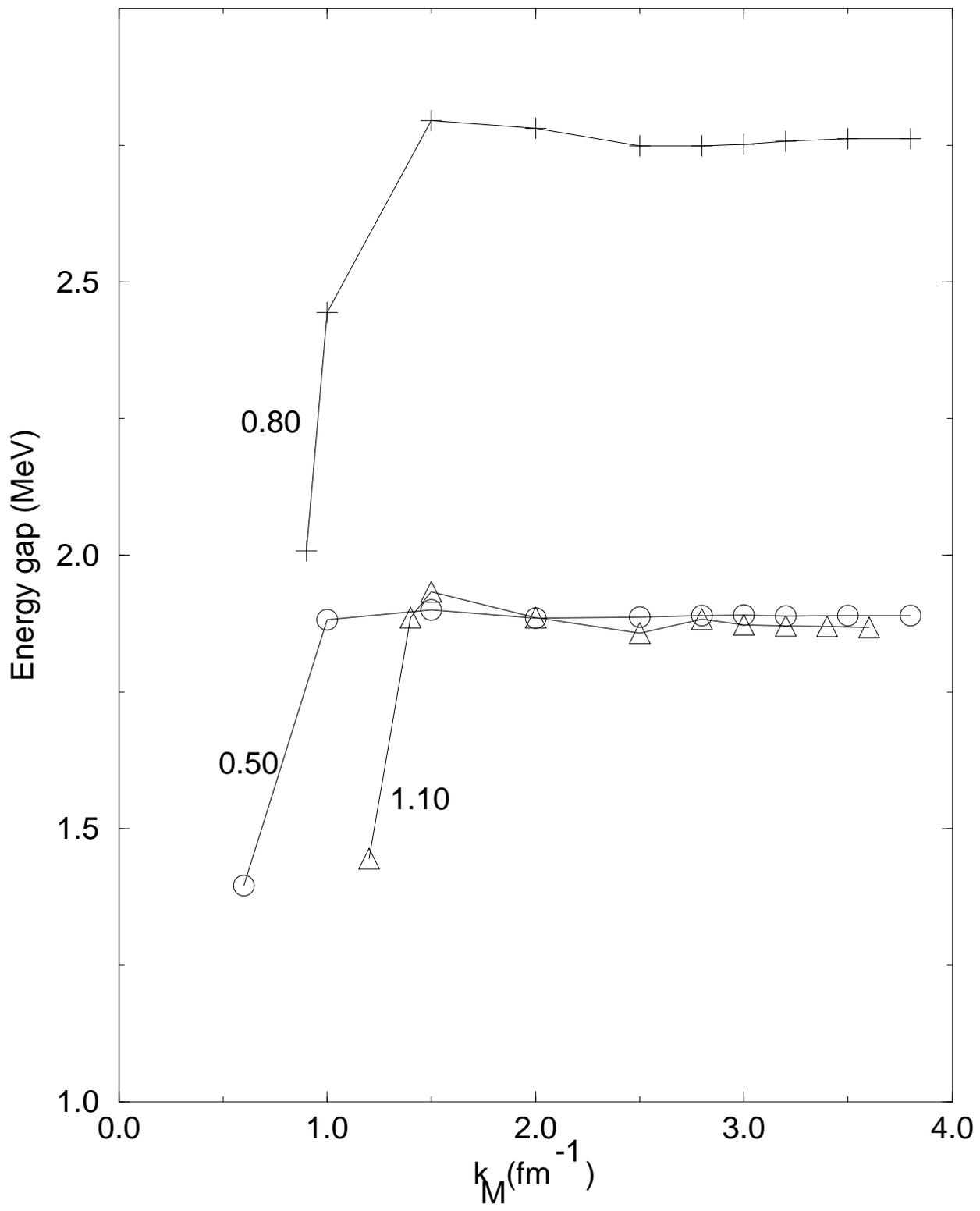

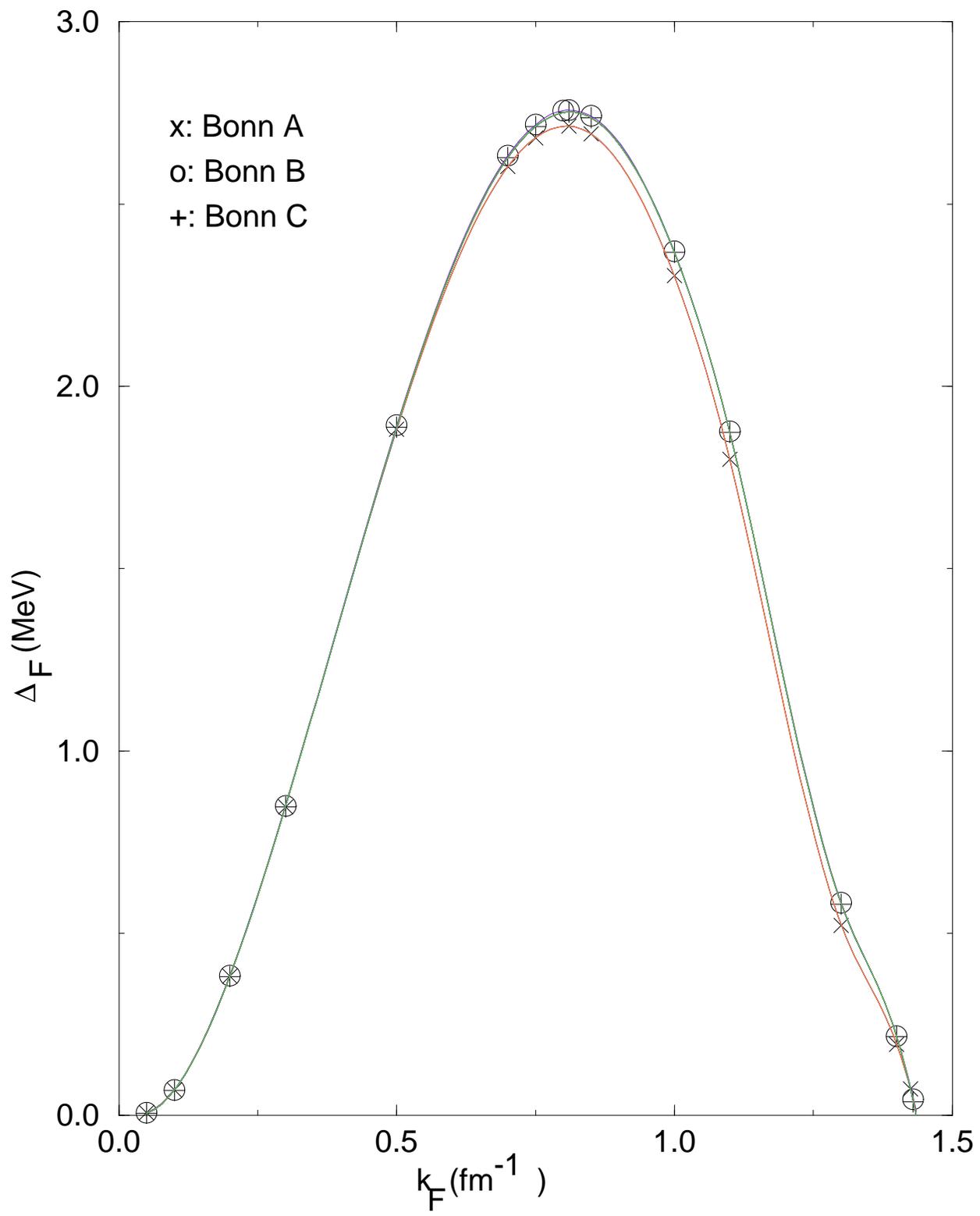

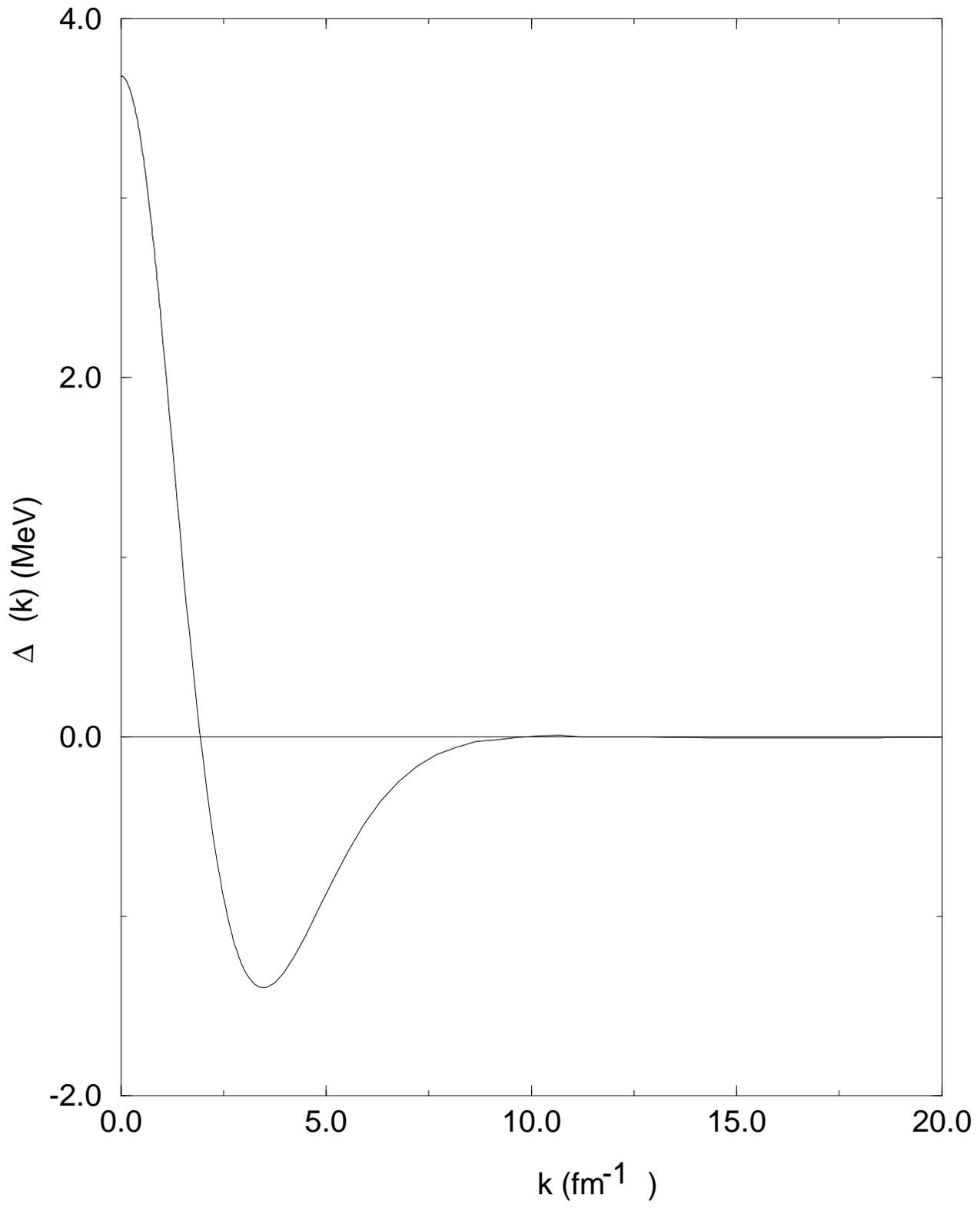

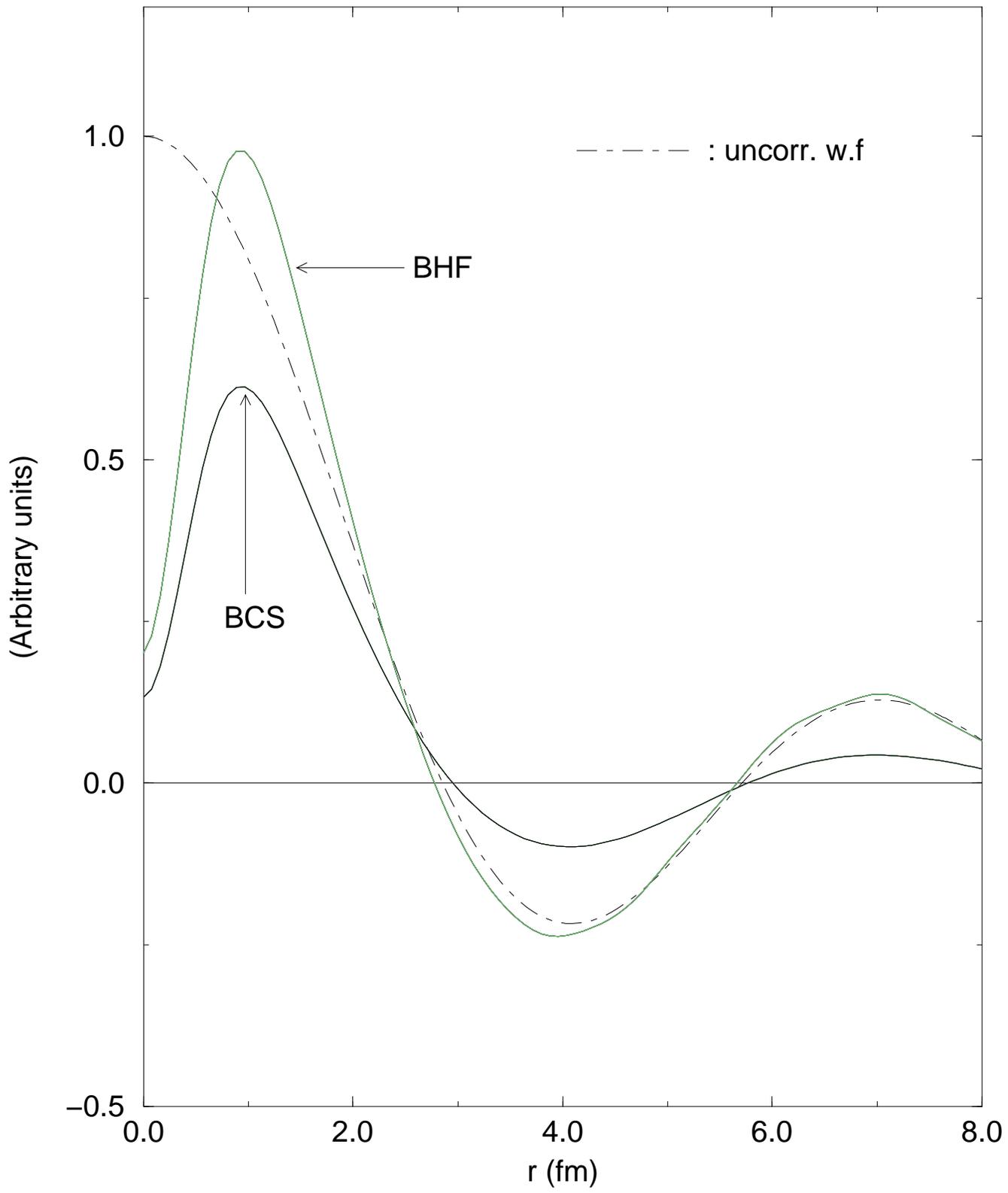

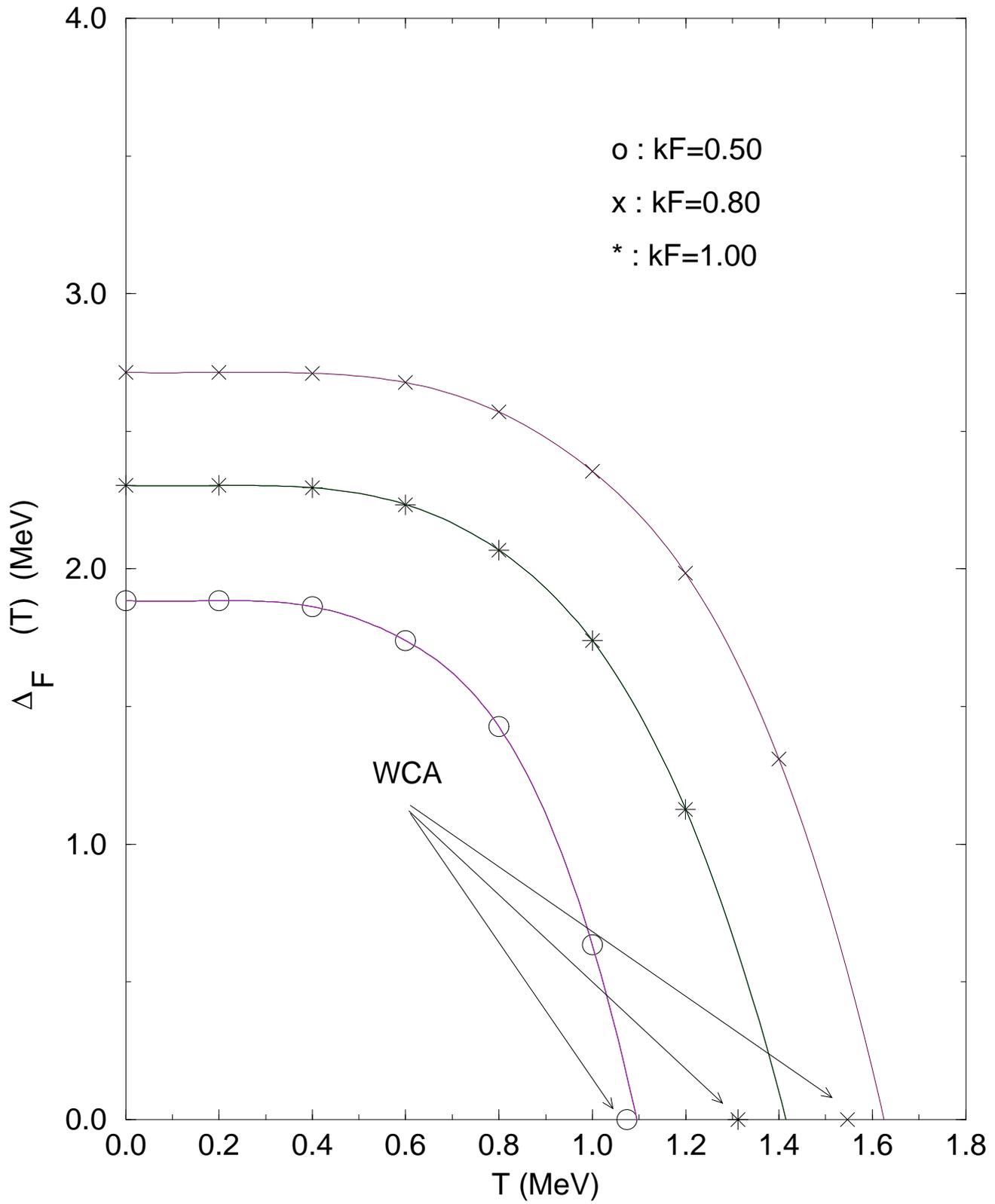

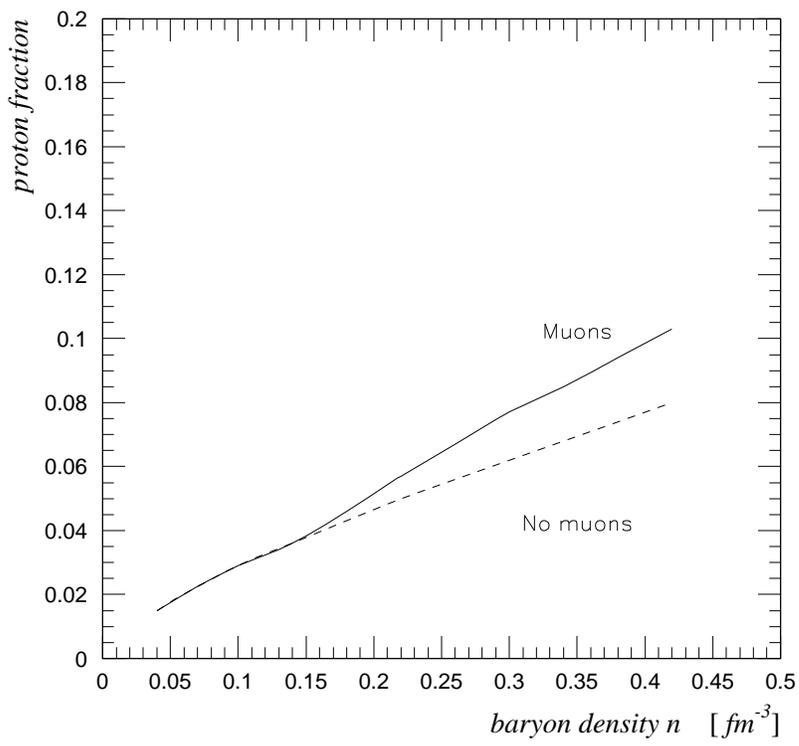

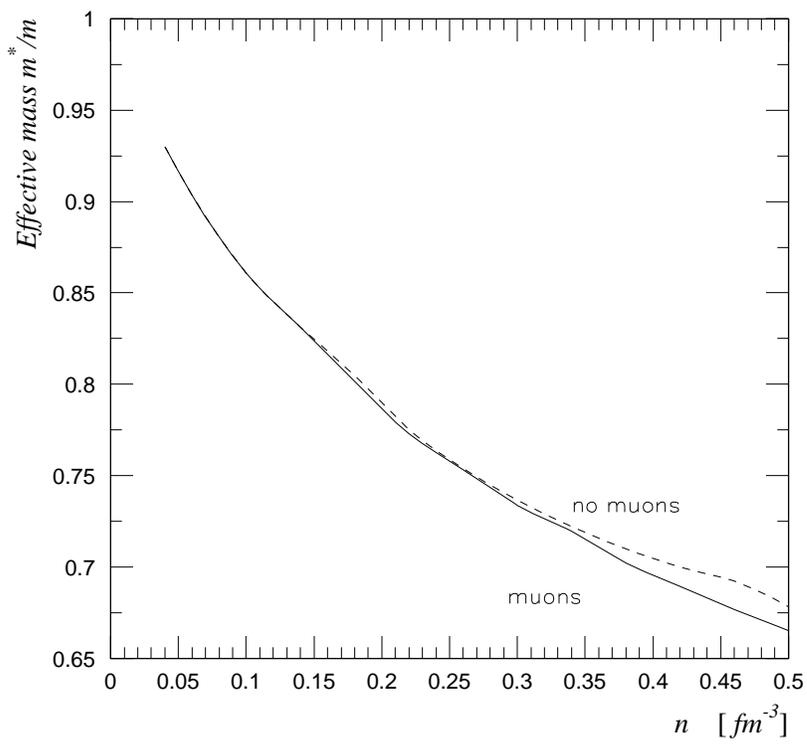

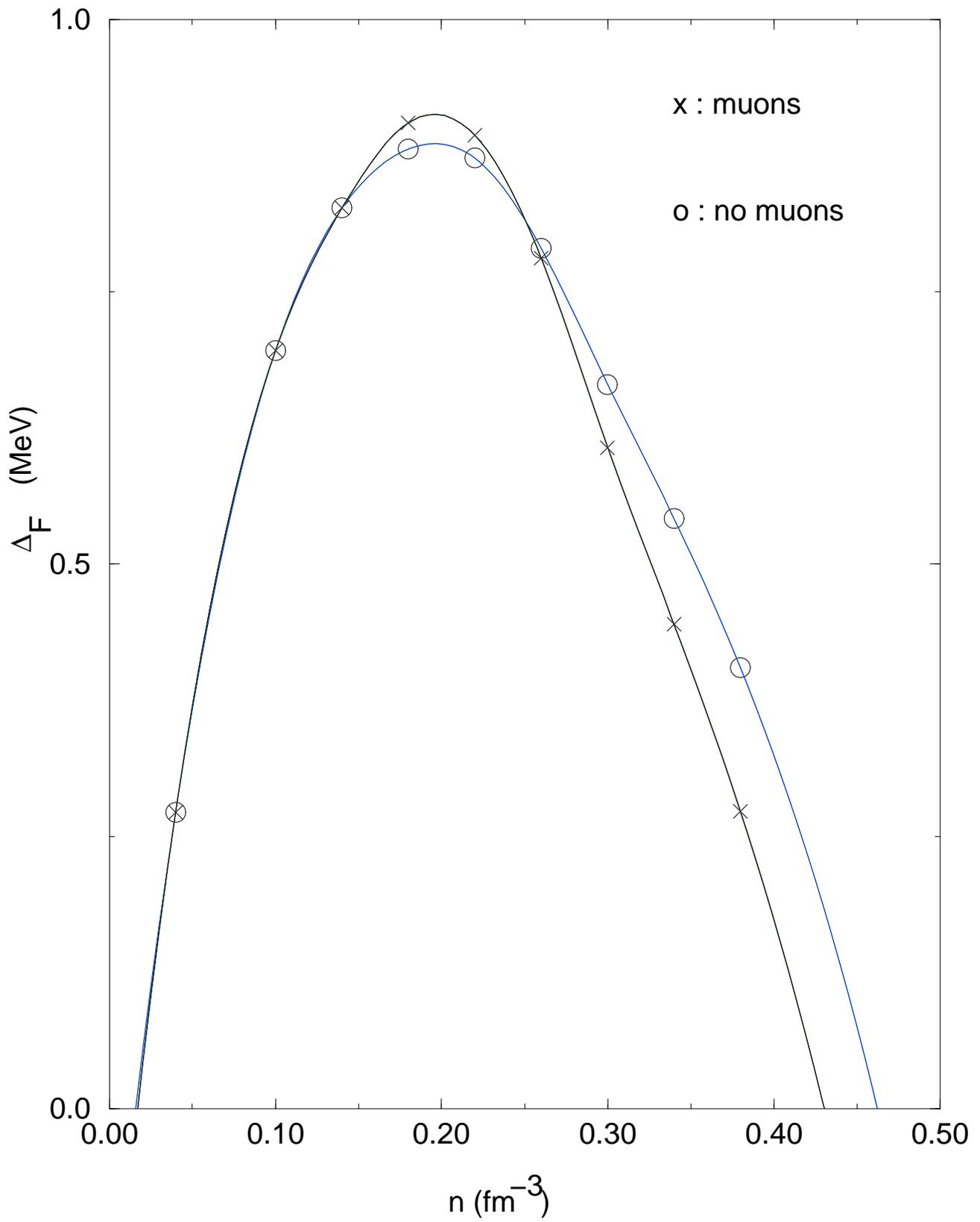

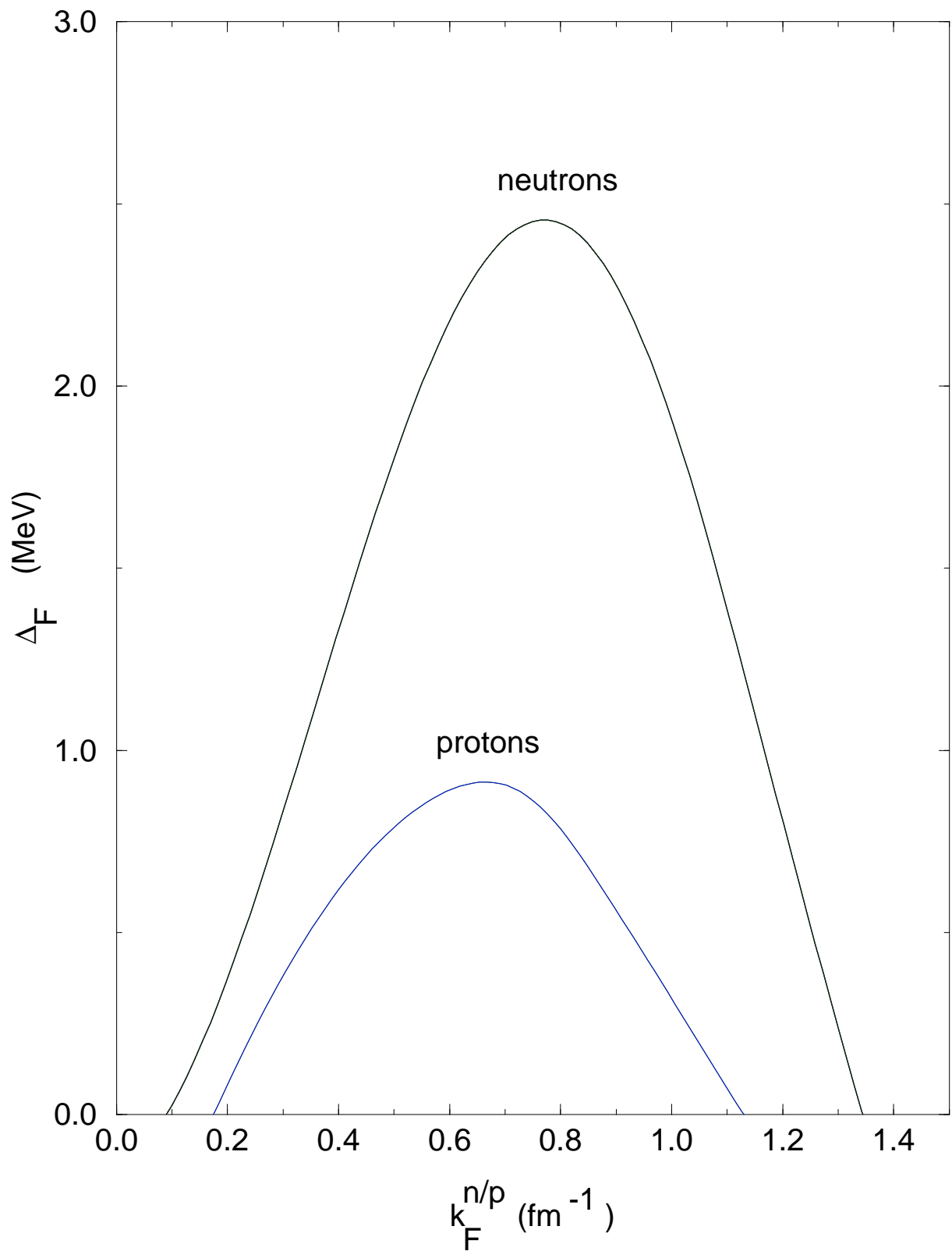

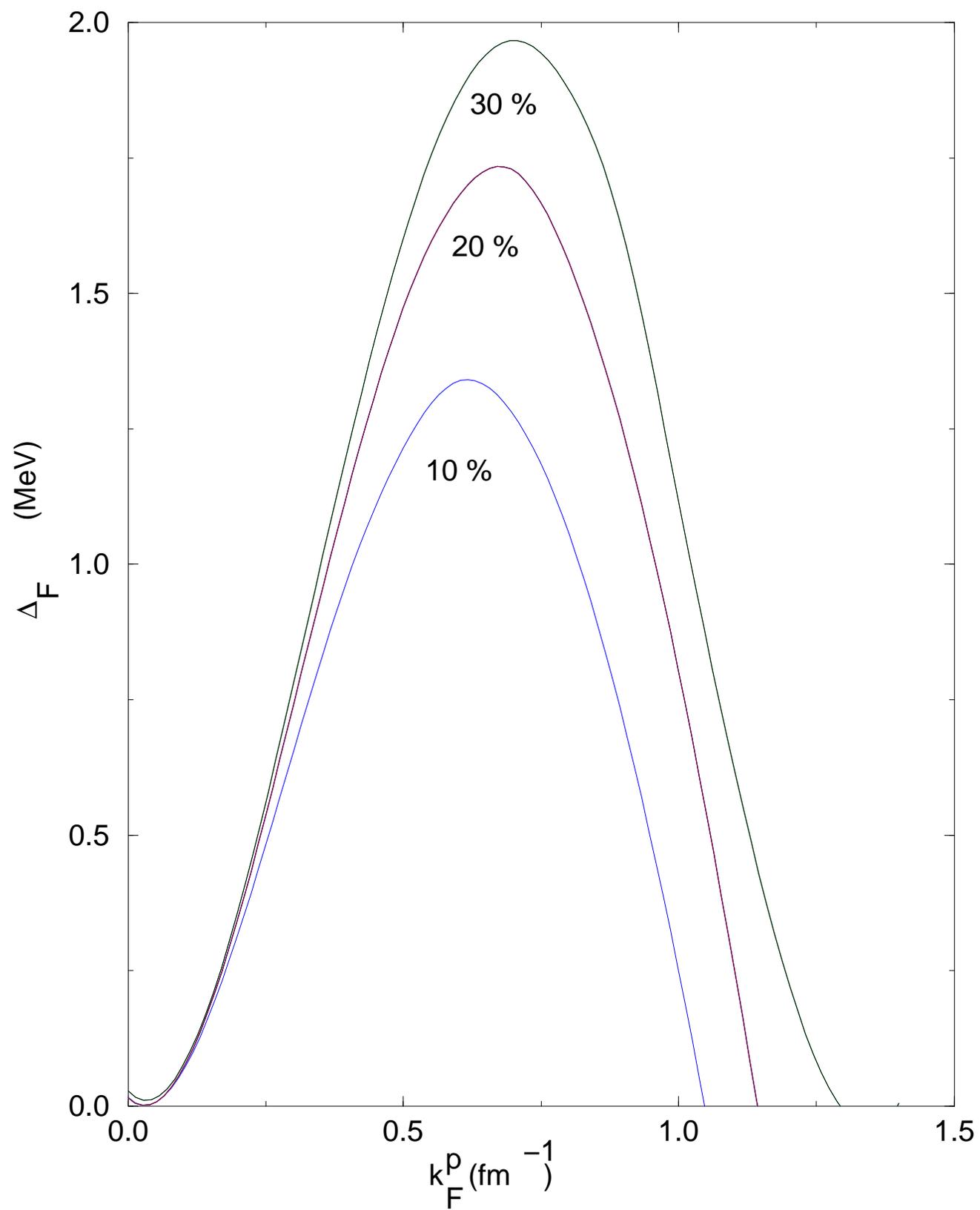

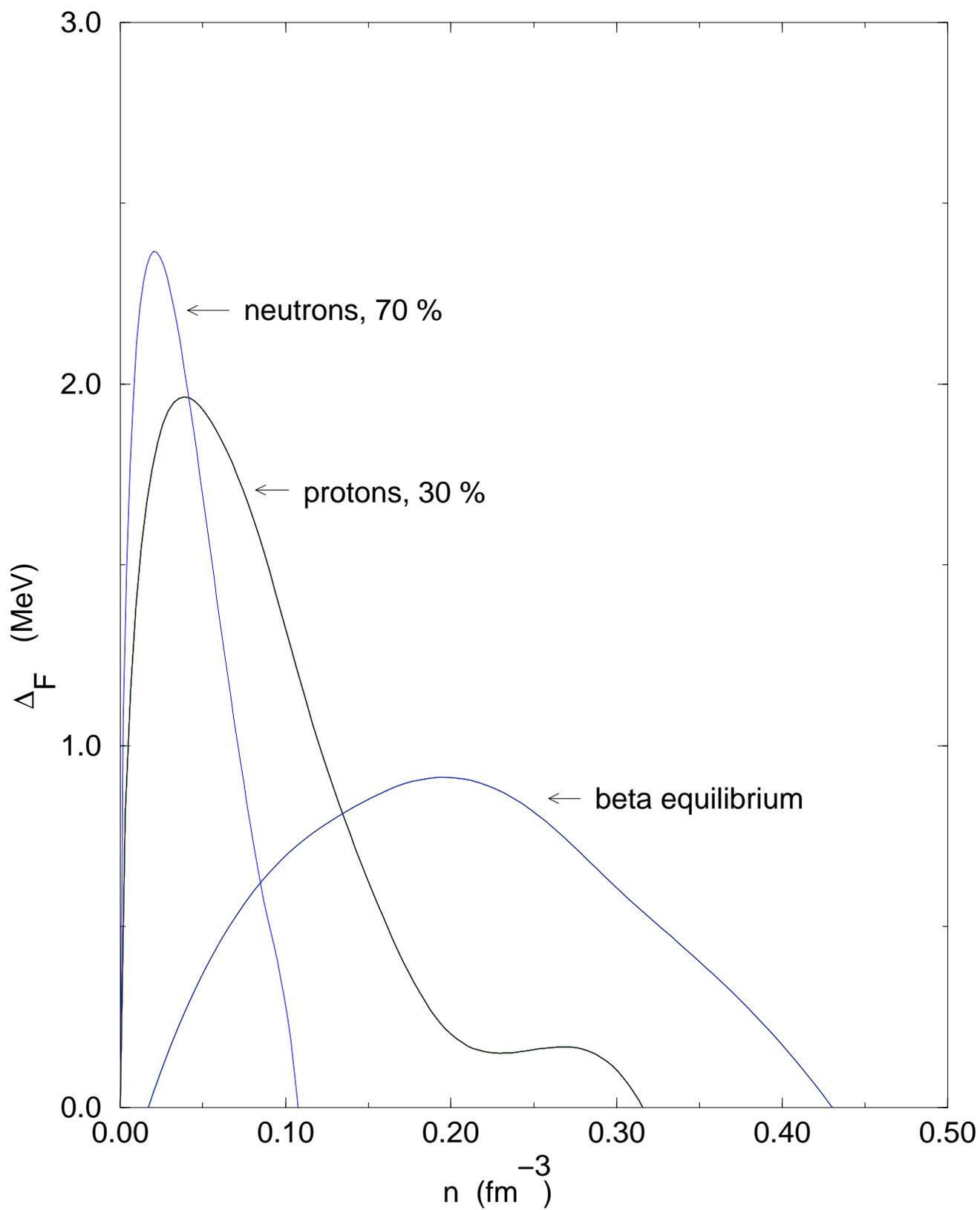